\documentclass[final,3p,authoryear,12pt]{elsarticle}
\usepackage[round]{natbib}

\usepackage[colorinlistoftodos,textsize=footnotesize]{todonotes}

\usepackage{amsfonts,amsmath,amssymb,lscape,float,graphicx,tabularx,url,times,theorem,color,natbib,afterpage,rotating,hyperref,xcolor,tikz,moresize,multirow,enumerate,colortbl,mathtools,bm,bbm}
\usepackage[utf8]{inputenc}
\usepackage{mathtools}
\usepackage[caption=true]{subfig}
\usepackage[ruled, vlined]{algorithm2e}
\usepackage{enumitem}

\mathtoolsset{showonlyrefs}
\allowdisplaybreaks 

\usepackage{diagbox}
\usepackage{bbm}

\usepackage{titlesec}
\titleformat{\subsection}    
       {\fontsize{12}{17}\bfseries}{\thesubsection}{1em}{}
\titleformat{\subsubsection}    
       {\fontsize{12}{17}\bfseries\itshape}{\thesubsubsection}{1em}{}
       
\titleformat{\paragraph}[runin]{\normalfont\normalsize}{\theparagraph}{1em}{}[\,\bfseries .]
\titlespacing*{\paragraph}{0pt}{3.25ex plus 1ex minus .2ex}{10pt}

\newcounter{algsubstate}

\usepackage{setspace} 
 \newcommand{\acknowledgements}[1]{%
  \nonumnote{#1}}

\definecolor{myblue}{rgb}{0.8,0.8,1}
\definecolor{myred}{rgb}{1,0.8,0.8}
\definecolor{mygreen}{rgb}{0.8,1,0.8}
\definecolor{mygrey}{rgb}{220,220,220}

\usepackage[bottom]{footmisc}
\usepackage[scaled]{helvet}

\DeclareMathAlphabet{\xcal}{OMS}{cmsy}{m}{n}

\definecolor{dbblue}{RGB}{10,65,155}				
\definecolor{dbred}{RGB}{215,0,50}					
\definecolor{blue}{RGB}{0,113.9850,188.9550} 			
\definecolor{red}{RGB}{216.7500,82.8750,24.9900} 		
\definecolor{green}{RGB}{118.8300,171.8700,47.9400} 	
\definecolor{grey}{RGB}{110,110,110}				
\definecolor{lgrey}{RGB}{210,210,210}				
\definecolor{c1}{RGB}{0,113.9850,188.9550}			
\definecolor{c2}{RGB}{216.7500,82.8750,24.9900}		
\definecolor{c3}{RGB}{236.8950,176.9700,31.8750}		
\definecolor{c4}{RGB}{125.9700,46.9200,141.7800}		
\definecolor{c5}{RGB}{118.8300,171.8700,47.9400}		
\definecolor{c6}{RGB}{76.7550,189.9750,237.9150}		
\definecolor{c7}{RGB}{161.9250,19.8900,46.9200}		
\definecolor{c14}{RGB}{0,102,102}					

\hypersetup{colorlinks,urlcolor=dbblue,citebordercolor=dbblue,linkbordercolor=dbblue,filecolor=dbblue,filebordercolor=dbblue,citecolor=dbblue,linkcolor=dbblue,colorlinks=true}

\defcitealias{UK7year:2013}{Griffiths et al., 2013}

\makeatletter
\def\ps@pprintTitle{%
  \let\@oddhead\@empty
  \let\@evenhead\@empty
  \def\@oddfoot{\reset@font\hfil\thepage\hfil}
  \let\@evenfoot\@oddfoot
}
\makeatother

\usepackage{graphicx}
\graphicspath{{images/}{../images/}}
\usepackage{enumerate}
\usepackage{bbold}
\usepackage{bbm}
\usepackage{subfig}
\usepackage{verbatim}
\usepackage{accents}

\newcommand{\ut}[1]{\underaccent{\tilde}{#1}}

\newcommand{\utilde}[1]{\ut{#1}}

\newtheorem{thm}{Theorem}
\newtheorem{crl}{Corollary}
\newtheorem{prop}{Proposition}

\newtheorem{assume}{Assumption}

\usepackage{geometry}
\begin{document}

\title{\textbf{Solvability of Differential Riccati Equations and Applications to Algorithmic Trading with Signals}}
\author{Fayçal Drissi}
\ead{faycal.drissi@eng.ox.ac.uk}

\address{Oxford-Man Institute of Quantitative Finance, University of Oxford}

\date{}

\journal{TBC}
\acknowledgements{The author is grateful to Álvaro Cartea, Patrick Chang, Olivier Guéant, Anthony Ledford, Leandro Sánchez-Betancourt, and the Oxford Victoria Seminar participants at Oxford for insightful comments. The author is grateful to the Oxford-Man Institute's generosity and hospitality.}


\begin{frontmatter}

\begin{abstract}
We study a differential Riccati equation (DRE) with indefinite matrix coefficients, which arises in a wide class of practical problems. We show that the DRE solves an associated control problem, which is key to provide existence and uniqueness of a solution. As an application, we solve two algorithmic trading problems in which the agent adopts a constant absolute risk-aversion (CARA) utility function, and where the optimal strategies use signals and past observations of prices to improve their performance. First, we derive a multi-asset market making strategy in over-the-counter markets, where the market maker uses an external trading venue to hedge risk. Second, we derive an optimal trading strategy that uses prices and signals to learn the drift in the asset prices.
\end{abstract}
 

\end{frontmatter}

\section{Introduction}

Matrix differential Riccati equations (DREs) with indefinite matrix coefficients, i.e., matrices that are neither positive nor negative semi-definite, arise in a wide class of control problems such as linear-quadratic zero-sum games, $\mathcal H^\infty$ control, linear-quadratic Gaussian (LQG) control problems with indefinite cost matrix, and linear-exponential-quadratic Gaussian (LEQG) control problems. 

Most results on these equations provide a statement of equivalence between the existence of solutions to a DRE and the solvability of a control problem. In this paper, we provide an existence and uniqueness result for the solution to a type of DRE with indefinite matrix coefficients which arises in various control setups. Special cases of the DRE are relevant for algorithmic trading problems in LEQG setups, where the cost is an exponential of a quadratic functional of the state and the controls. In particular, the quadratic term matrix coefficient has both positive and negative eigenvalues, so classical existence results are not readily available.



We use the result to provide solutions to two algorithmic portfolio trading problems; one is market making with signals and hedging, the second is optimal trading with Bayesian learning of the drift. In both problems, the optimal strategies take the form of a linear feedback control whose feedback gain is obtained from the solution to a special case of the DRE that we study. Efficient numerical solving techniques for such equations have been extensively studied and provide computationally efficient methods to market participants, for whom speed and precision are key for better performance.

\paragraph{\textbf{Market making}}  As a first application for liquidity providers, we introduce a comprehensive framework for multi-asset market making in OTC markets with three important features : $(i)$ portfolio dynamics that incorporate signals, $(ii)$ external hedging and client tiering, and $(iii)$ computationally efficient methods to obtain quotes for large portfolios.

OTC markets are off-exchange and quote-driven markets based on a network of market makers that allow clients to buy and sell securities. In these markets, quotes are either streamed to clients or clients request for quotes (RFQs).\footnote{For a better understanding of the RFQ system, we refer to \cite{fermanian2016behavior}.} In practice, OTC market makers provide liquidity for assets within a specific sector or asset class. Thus, the price dynamics of assets for which they make quotes usually share common stochastic trends and co-movements. For example, in OTC bond markets, a large number of bonds from different issuers are traded and one needs to account for the joint dynamics that they exhibit; appropriate dynamics are necessary to obtain quotes that balance risk at the portfolio level. 

Furthermore, market makers often use price predictors to improve trading performance, and use instruments that offset their inventory risk to hedge their positions. Our framework uses multivariate Ornstein--Uhlenbeck (multi-OU) dynamics to model the joint dynamics of these variables. We use an approximation technique similar to that introduced in \cite{bergault2018closed} and we solve the optimal quote and optimal hedging approximation problem using the DRE existence and uniqueness result. Finally, we study the quoting strategy with numerical examples for a market maker in charge of a single asset with mean reverting prices, and for a market maker in charge of a pair of cointegrated assets.

\paragraph{\textbf{Optimal trading}} As a second application for liquidity takers, we combine the tools of stochastic control and online learning to derive an optimal portfolio trading strategy. The agent uses observed prices and signals to continuously update her estimate of the drift in the prices. The strategy uses signals either to improve the performance of execution programmes or to execute statistical arbitrages. We show that the problem reduces to a DRE for which existence and uniqueness of a solution is a straightforward application of our result. 

In most execution strategies in the literature, model parameters are estimated and fixed prior to the start of the trading window. There are two downsides to this approach. First, parameter misspecification makes the strategy non-optimal. Second, the strategy does not incorporate uncertainty about the parameters. This work builds on the work in \cite{bismuth2019portfolio} and proposes a model where the agent uses a prior distribution over the drift parameter to encode her uncertainty, and continuously uses price and signal observations to update her estimate during the execution programme.

\paragraph{\textbf{Literature review}} 

Classical results on LQG problems and the associated DREs are in \cite{abou2012matrix, fleming2012deterministic}. \cite{jacobson1973optimal} uses optimal control to solve an LEQG control problem and obtains optimal controls that are equivalent to those obtained for quadratic differential games. \cite{whittle1990risk} and  \cite{lim2005new} use the stochastic maximum principle to obtain similar results. Our existence and uniqueness result generalises that of \cite{bergault2021multi}; in particular, the matrix coefficients are deterministic time-dependent matrix functions and the quadratic term in the cost functional depends on the observed system. The existence and uniqueness of solutions in general remains an open problem for matrix DREs with indefinite matrix coefficients. This work relates to the literature  that addresses the solvability of these equations in special cases; see \cite{chen1998stochastic, rami2001solvability, hu2003indefinite}. Finally, matrix DREs arise in various algorithmic trading problems; see \cite{cartea2019trading, cartea2022optimalElec, barzykin2022dealing}, 

In market making, the first approaches are in \cite{ho1983dynamics, glosten1985bid, avellaneda2008high}. These frameworks have been extended in many directions; see \cite{gueant2012optimal, gueant2013dealing, cartea2014buy}, and the books \cite{cartea2015algorithmic, gueant2016financial}. Our work is related to the literature addressing the curse of dimensionality in multi-asset market making; see \cite{gueant2013dealing, gueant2017optimal, gueant2019deep}. A strand of the algorithmic trading literature uses market signals to improve the performance of strategies; see \cite{cartea2016incorporating, cartea2020market, forde2022optimal, fouque2021optimal, cartea2022decentralised, cartea2022decentralised2}.

In optimal execution, the first approaches are in \cite{bertsimas1998optimal, almgren1999value}, and model extensions are in \cite{schied2009risk, alfonsi2010optimal, gatheral2012transient, forsyth2012optimal,  gueant2015optimal, cartea2021latency, cartea2022optimal}; see the comprehensive review \cite{donnelly2022optimal}. Several papers generalise the classical approaches of optimal execution to more complex price dynamics; see \cite{cartea2016incorporating, almgren2012optimal}. In particular, similar to this work, \cite{cartea2019trading} and \cite{bergault2021multi} study optimal trading models where the joint dynamics of prices follow a multi-OU process. Our work also relates to the literature which incorporates uncertainty and learning in execution. \cite{laruelle2013optimal} derive a strategy where the agents learns the parameters of a jump process, \cite{cartea2017algorithmic} incorporate model uncertainty in execution, and \cite{casgrain2019trading} derive strategies with learning of latent state distribution upon which prices depend. Similar to the problem addressed in this work, \cite{bhudisaksang2021adaptive} introduce an adaptive control framework which is robust to model misspecification, where the agent learns the drift with jump-diffusion uncertainty. Recently, \cite{cartea2022brokers} derive closed-form strategies where a broker uses the toxic flow to extract alpha signals, and \cite{cartea2023bandits} use Gaussian processes to learn the mapping from signal values to trading performance.


The remainder of this paper proceeds as follows. Section \ref{sec:riccati} introduces a family of DREs with indefinite matrix coefficients and provides existence and uniqueness results based on a set of assumptions. Section~\ref{sec:begvou} derives a portfolio market making strategy that uses signals to improve trading performance and uses a trading venue to hedge risk. The joint dynamics of the assets in the portfolio, the hedging instruments, and the signals follow multi-OU dynamics. We show that the approximation of the optimal quoting and hedging strategies reduces to a DRE that is a special case of that studied in  Section \ref{sec:riccati}. Finally, Section \ref{sec:bayesien} solves optimal portfolio trading with Bayesian learning of the drift, using signals and past observations of the price to improve performance. We show that the problem reduces to a DRE that is also a special case of that studied in  Section \ref{sec:riccati}.  

\section{A general differential matrix Riccati equation \label{sec:riccati}}

DREs with indefinite matrix coefficients arise in LQ games \citep[][]{jacobson1973optimal}, $\mathcal H^\infty$ control \citep[][]{limebeer1992game}, LQG problems with indefinite cost \citep[][]{chen1998stochastic}, and LEQG problems \citep[][]{whittle1990risk}. Their solvability is also key in practical financial decision problems, such as portfolio optimisation \citep[][]{zhou2000continuous}, option pricing \citep[][]{kohlmann2000relationship}, and algorithmic trading \citep[][]{bergault2021multi}. The results in these works  show that some control problem is well-posed if there are solutions to a DRE from which one obtains the optimal feedback control. In this section, we study a special type of DREs and provide conditions for the existence and uniqueness of a solution.

We motivate the DRE that we study by the type of DREs that arise in algorithmic trading problems where an agent adopts a CARA utility and where the dynamics of the asset prices are (linearly) driven by signals and past observations of prices. Section \ref{sec:begvou} and Section \ref{sec:bayesien} give two examples of such problems.

In the remainder of this paper, the symbol $\dot P$ denotes the first derivative of $P$, the superscript $^\intercal$ stands for the transpose operator, and $\mathbb N^\star$ is the set of positive integers. $\mathcal M_{d,r}(\mathbb R)$ is the set of $d\times r$ real matrices and $\mathcal M_d (\mathbb R) := \mathcal M_{d,d}(\mathbb R)$ is the set of $d \times d$ real square matrices. We also denote the set of real symmetric $d \times d$ matrices by $\mathcal S_d (\mathbb R)$. Finally, the subset of positive-definite and positive semi-definite matrices of $\mathcal S_d (\mathbb R)$ are respectively denoted by $\mathcal S^{++}_d (\mathbb R)$ and $\mathcal S^{+}_d (\mathbb R)$.

We study the matrix DRE
\begin{align}
\label{eq:RiccatiGeneral}
\dot{P}\left(t\right)=&\, Q\left(t\right)+Y\left(t\right)^{\intercal}P\left(t\right)+P\left(t\right)Y\left(t\right)+P\left(t\right)U\left(t\right)P\left(t\right), \quad \forall t\in[0,T],
\end{align}
with terminal condition $P\left(T\right),$ where $P \in \mathcal S_{d+r}\left(\mathbb R\right),$  $r\le d,$ and $(r,d)\in\mathbb N^\star \times \mathbb N^\star$. Let $\rho>0$, the matrix coefficients $Q \in \mathcal{C}^{0}\left(\left[0,T\right],\mathcal{S}_{d+r}(\mathbb{R})\right),$ $Y \in \mathcal{C}^{0}\left(\left[0,T\right],\mathcal{M}_{d+r}(\mathbb{R})\right),$ and $U \in \mathcal{C}^{0}\left(\left[0,T\right],\mathcal{S}_{d+r}(\mathbb{R})\right)$ are
\begin{align}
Q\left(t\right)=\left(\begin{matrix}
Q_{11}\left(t\right) & 0\\
0 & 0
\end{matrix}\right),\quad Y\left(t\right)=\left(\begin{matrix}
Y_{11}\left(t\right) & 0\\
Y_{21}\left(t\right) & 0
\end{matrix}\right),\quad U\left(t\right)=\left(\begin{matrix}
\rho\,U_{11}\left(t\right) & 0\\
0 & -U_{22}\left(t\right)
\end{matrix}\right),
\end{align}
and the terminal condition is 
\begin{align}
\label{eq:termcondP}
P\left(T\right)=- \left(\begin{matrix}
\varPsi & \frac{1}{2}\varUpsilon\\
\frac{1}{2}\varUpsilon^{\intercal} & \Gamma
\end{matrix}\right),
\end{align}
where $\ \ Q_{11}\in\mathcal{C}^{0}\left(\left[0,T\right],\ \mathcal{S}_{d}(\mathbb{R})\right),\ \ Y_{11}\in\mathcal{C}^{0}\left(\left[0,T\right],\ \mathcal{M}_{d}(\mathbb{R})\right),\ \ Y_{21} \in\mathcal{C}^{0}\left(\left[0,T\right],\ \mathcal{M}_{r,d}(\mathbb{R})\right),\ \ $  $  U_{11}\in\mathcal{C}^{0}\left(\left[0,T\right],\mathcal{S}_{d}^{+}(\mathbb{R})\right),$ $\ \ U_{22}\in\mathcal{C}^{0}\left(\left[0,T\right],\mathcal{S}_{r}^{++}(\mathbb{R})\right),$ $\ \ \varPsi\in\mathcal{S}_{d}^{+}(\mathbb{R}),$ $\ \ \Gamma\in\mathcal{S}_{r}(\mathbb{R}),\,$ and $\ \ \varUpsilon\in\mathcal{M}_{d,r}(\mathbb{R})\,.$ \\


We show that a solution  to the DRE \eqref{eq:RiccatiGeneral} with terminal condition \eqref{eq:termcondP} exists and is unique when the coefficients satisfy a set of assumptions. First, Subsection \ref{sec:LQGexp} shows that the solution to the DRE corresponds to the solution to an LEQG control problem. Next, Subsection \ref{sec:subsec:existence} uses the control problem and our assumptions to provide a-priori bounds to the solution to the DRE, and subsequently obtains existence and uniqueness of a solution.

\subsection{An equivalent exponential utility control problem \label{sec:LQGexp}}

Here, we show that the DRE \eqref{eq:RiccatiGeneral} corresponds to an LEQG control problem, which then becomes the key tool to prove the main result of existence and uniqueness of a solution in Subsection~\ref{sec:subsec:existence}. Let $\left(\Omega, \mathcal F, \mathbb P; \mathbb F = (\mathcal F(t))_{t \in [0,T]} \right)$ be a filtered probability space that satisfies the usual conditions and which supports all the processes defined in this section. Consider the following deterministic matrix functions:
\begin{align}
\label{eq:matrixvarchange}
\begin{cases}
A\left(t\right)= & U_{22}\left(t\right)^{-1}\,,\\
B\left(t\right)= & 2\,U_{22}\left(t\right)Y_{21}\left(t\right)\,,\\
C\left(t\right)= & Q_{11}\left(t\right)+Y_{21}\left(t\right)^{\intercal}U_{22}\left(t\right)Y_{21}\left(t\right)\,,
\end{cases}\quad\text{and}\quad\begin{cases}
R\left(t\right)= & -Y_{11}\left(t\right)\,,\\
\Sigma\left(t\right)= & \frac{1}{2}\,U_{11}\left(t\right)\,.
\end{cases}
\end{align}

Note that $\ A\in \mathcal C^0 \left([0,T],\ \mathcal S_r^{++}(\mathbb R \right)),\ B \in \mathcal C^0 \left([0,T],\, \mathcal M_{r,d}( \mathbb R) \right),\, C \in \mathcal C^0 \left([0,T], \mathcal S_d ( \mathbb R ) \right),$ $R \in \mathcal C^0 \left([0,T],\mathcal M_d ( \mathbb R ) \right),\,$ and~$\, \Sigma \in \mathcal C^0 \left([0,T],\, \mathcal S_d^+( \mathbb R) \right).$ Furthermore, let $j \in \mathbb{N}^\star$ be the rank of the matrix $\Sigma(t)$ and consider a matrix $V \in \mathcal C^0 \left([0,T], \, \mathcal M_{d,j}(\mathbb R)\right)$ such that $\Sigma\left(t\right) = V\left(t\right) V\left(t\right)^\intercal $ for all $t \in [0,T].$ The change of variables in \eqref{eq:matrixvarchange} leads to the following matrix coefficients for the DRE \eqref{eq:RiccatiGeneral}:
\begin{align}
Q(t)=&\left(\begin{matrix}
C(t)-\frac{1}{4}B(t)^{\intercal}A(t)\,B(t) & 0\\
0 & 0
\end{matrix}\right), \ \, Y(t)=\left(\begin{matrix}
-R\left(t\right) & 0\\
\frac{-1}{2}\,A(t)\,B(t) & 0
\end{matrix}\right), \ \, U(t)=\left(\begin{matrix}
2\,\rho\Sigma(t) & 0\\
0 & -A(t)^{-1}
\end{matrix}\right)\,.
\end{align}

Consider an observable system described by a $d$-dimensional state processes $\left(x_t\right)_{t\in[0,T]}$ with linear dynamics 
\begin{align} \label{eq:gen_dynx}
dx_{t}&=R\left(t\right)\,x_{t}\,dt+V\left(t\right)dW_{t},
\end{align}
where $x_0 \in \mathbb R^d$ is known and $T>0$ is fixed. The matrix function $R$ encodes the linear dependence of the system to its previous state, $V$ scales the variance of the noise, and $(W_t)_{t \in [0,T]} = \left(W^1_t, \ldots, W^j_t \right)^\intercal_{t \in [0,T]}$ is a $j$-dimensional standard Brownian motion with independent coordinates. Note that the SDE \eqref{eq:gen_dynx} admits a strong solution; see \ref{appendixproofprop1}.

An agent controls an $r$-dimensional process  $(u_t)_{t \in [0,T]}$ that affects two measured outputs $(y_t)_{t \in [0,T]}$ and  $(z_t)_{t \in [0,T]}.$ The set of admissible controls is 
\begin{align}\label{def:AdmissibleSet_t}
\mathcal A_t = \left\{ (u_s)^\intercal_{s \in [t,T]},\ \mathbb R^r\textrm{-valued},\, \mathbb F\textrm{-adapted, satisfying a linear growth condition w.r.t } (x_s)_{s \in [t,T]}  \right\},
\end{align}
where we define the \emph{linear growth condition} on $[t,T]$ with respect to $(x_s)_{s \in [t,T]}$ to be satisfied by an $\mathbb R^r$-valued, $\mathbb F$-adapted process $(\zeta_s)_{s \in [t,T]}$ if there exists a constant $C_{t,T}>0$ such that for all $s \in [t,T]$,
$\| \zeta_s \| \le C_{t,T} \left(1 + \underset{\tau \in [t,s]}{\sup} \| x_\tau \| \right)$ almost surely.\footnote{Here, $\|\cdot\|$ is a fixed norm on $\mathbb R^d.$} We write $\mathcal A := \mathcal A_0.$

The output $y$ is affected linearly and follows the dynamics 
\begin{align} \label{eq:gen_dyny}
dy_{t}&=u_{t}\,dt,
\end{align}
where $y_0 \in \mathbb R^r$ is known. On the other hand, the output $z$ takes a quadratic form in the system and the controls and follows the dynamics  
\begin{align}
\label{eq:gen_dynz}
dz_{t}&=-\left(\begin{matrix}
u_{t}\\
x_{t}
\end{matrix}\right)^{\intercal}\left(\begin{matrix}
A\left(t\right) & \frac{1}{2}B\left(t\right)\\
\frac{1}{2}B\left(t\right)^{\intercal} & C\left(t\right)
\end{matrix}\right)\left(\begin{matrix}
u_{t}\\
x_{t}
\end{matrix}\right)dt,
\end{align}
where $z_0 \in \mathbb R$ is known. Note that the condition $r\le d$ above guarantees that the dimension of the agent's control $u$ is smaller than that of the system $x$. In our problem, the output $y$ measures an objective that the agents targets at the terminal time $T$, and the output $z$ measures an objective that the agent wishes to maximise or a cost that she wishes to minimise by the terminal time $T.$  

The performance criterion of the agent associated with a control $u \in \mathcal A,$ is   \begin{align}
\label{eq:perfcriterion0}
\mathbb{E}\left[-\exp\left(-\rho\left(z_{T}-\left(\begin{matrix}
x_{T}\\
y_{T}
\end{matrix}\right)^{\intercal}\left(\begin{matrix}
\varPsi & \frac{1}{2}\varUpsilon\\
\frac{1}{2}\varUpsilon^{\intercal} & \Gamma
\end{matrix}\right)\left(\begin{matrix}
x_{T}\\
y_{T}
\end{matrix}\right)\right)\right)\right],
\end{align}
where $\rho$ measures risk aversion. 

We introduce the following assumptions that are  subsequently used in the remainder of this section: 
\begin{assume}\label{assume:1}
$ $ \begin{enumerate}[label=(\roman*)]
    \item $C\left(t\right) = Q_{11}\left(t\right)+Y_{21}\left(t\right)^{\intercal}U_{22}\left(t\right)Y_{21}\left(t\right) \in \mathcal S_d^{+}\left(\mathbbm R\right)\ $ for  all $\, t\in[0,T]\,,$ \label{assumptions:1:i}
    \item $B(t)\,\Sigma(t)\,B(t)^\intercal \in \mathcal S_d^{++}\left(\mathbbm R\right), \ \,$ for all $\, t\in[0,T]\,,$
    \label{assumptions:1:ii}
    \item $Y_{11}\left(s\right)\ $ and $\ Y_{11}\left(t\right)$ commute for all $\, \left(s, t\right) \in [0,T] \times [0,T]\,.$
    \label{assumptions:1:iii}
\end{enumerate}

\end{assume}

\begin{assume}\label{assume:2}
$\, \forall t\in[0,T], \ C(t) = \varPsi = 0$.
\end{assume}

In the optimisation problem \eqref{eq:perfcriterion0}, Assumption \ref{assume:1}-\ref{assumptions:1:i} ensures that the term in the performance criterion that is quadratic in the state $x$ is a cost and not a gain.  Assumptions  \ref{assume:1}-\ref{assumptions:1:ii}  and  \ref{assume:1}-\ref{assumptions:1:iii} are technical conditions to obtain an upper bound for the value function which results in an upper bound for the solution to the DRE; see \ref{appendixproofthm2}. Finally, when $r = d$, when the Assumptions  \ref{assume:1} and \ref{assume:2} hold, and when the matrix coefficients in \eqref{eq:RiccatiGeneral} are constant, the DRE \eqref{eq:RiccatiGeneral} is similar to that in \cite{bergault2021multi}.

The agent maximises her performance \eqref{eq:perfcriterion0} and her value function $\vartheta:[0,T] \times \mathbb R^d \times \mathbb R^r \times \mathbb R \rightarrow \mathbb R$ is 
\begin{align}
\label{eq:vfdef}
\vartheta(t,x,y,z) =\sup_{u\in\mathcal{A}_{t}}\mathbb{E}\left[-\exp\left(-\rho\left(z_{T}^{t,x,z,u}-\left(\begin{matrix}
x_{T}^{t,x}\\
y_{T}^{t,y,u}
\end{matrix}\right)^{\intercal}\left(\begin{matrix}
\varPsi & \frac{1}{2}\varUpsilon\\
\frac{1}{2}\varUpsilon^{\intercal} & \Gamma
\end{matrix}\right)\left(\begin{matrix}
x_{T}^{t,x}\\
y_{T}^{t,y,u}
\end{matrix}\right)\right)\right)\right],
\end{align}
where the processes $(x^{t,x}_s)_{s \in [t,T]},$ $(y^{t,y,u}_s)_{s \in [t,T]}$, and $(z^{t,x,z,u}_s)_{s \in [t,T]}$ follow the dynamics 
\begin{equation}
\label{eq:controllerdynamics}
dz_{s}^{t,x,z,u}=-\left(\begin{matrix}
u_{s}\\
x_{s}^{t,x}
\end{matrix}\right)^{\intercal}\left(\begin{matrix}
A\left(s\right) & \frac{1}{2}B\left(s\right)\\
\frac{1}{2}B\left(s\right)^{\intercal} & C\left(s\right)
\end{matrix}\right)\left(\begin{matrix}
u_{s}\\
x_{s}^{t,x}
\end{matrix}\right)ds \quad \text{and}\quad \begin{cases}
dx_{s}^{t,x} & =R\left(s\right)x_{s}^{t,x}ds+V\left(s\right)dW_{s},\\
dy_{s}^{t,u} & =u_{s}ds,
\end{cases}
\end{equation}
for $(t,x,y,z) \in [0,T] \times \mathbb R^d \times \mathbb R^r \times \mathbb R$ and $u \in \mathcal A_t$, with $x_t^{t,x} =x,$ $y_{t}^{t,y,u} = y$ and $z_{t}^{t,x,z,u} = z$. We state the following result for which a proof is given in \ref{appendixproofprop1}.
\begin{prop} \label{prop1}
Let Assumption \ref{assume:1} hold. If Assumption \ref{assume:2} also holds,  then $\vartheta(t,x,y,z) > -\infty\,.$ Otherwise, if Assumption \ref{assume:2} does not hold, then $\exists \, \overline \rho > 0$ such that $\forall \rho \in (0, \overline\rho], \ \vartheta(t,x,y,z) > -\infty\,.$ 
\end{prop}

The next theorem shows that the solution to the optimisation problem \eqref{eq:perfcriterion0} is associated with that of the DRE \eqref{eq:RiccatiGeneral}; for a proof see \ref{appendixproofthm1}.

\begin{thm} \label{thm1}
Assume $M_1 \in C^1 \left([0,T], \, \mathcal S_d(\mathbb R) \right)$, $M_2 \in C^1 \left([0,T], \,\mathcal M_{d,r}(\mathbb R) \right),$ and $M_3 \in C^1 \left([0,T], \, \mathcal S_r(\mathbb R) \right)$ satisfy the DRE \eqref{eq:RiccatiGeneral} in $$P\left(t\right)=\left(\begin{matrix}
M_{1}\left(t\right) & \frac{1}{2}M_{2}\left(t\right)\\
\frac{1}{2}M_{2}\left(t\right)^{\intercal} & M_{3}\left(t\right) \end{matrix}\right),$$ with terminal condition \eqref{eq:termcondP}, where the coefficients are in \eqref{eq:matrixvarchange}. Define $M_6 \in C^1 \left([0,T], \mathbb R \right)$ as  \begin{align}\label{eq:thm1:M6} M_{6}\left(t\right)= -\int_0^t \text{\normalfont Tr}\left(\Sigma\left(s\right)M_{1}\left(s\right)\right) ds,\end{align}
where $\text{\normalfont Tr}$ denotes the trace operator. 
Let the function $\theta:[0,T]\times \mathbb R^d\times \mathbb R^r \mapsto \mathbb R$ and the associated function $w:[0,T]\times\mathbb R^d\times \mathbb R^r\times \mathbb R \mapsto \mathbb R$ be 
\begin{align} \label{eq:ansatz1}
\theta\left(t,x,y\right)=x^{\intercal}M_{1}\left(t\right)x+x^{\intercal}M_{2}\left(t\right)y+y^{\intercal}M_{3}\left(t\right)y+M_{6}\left(t\right),
\end{align}  \begin{align} \label{eq:ansatz2}
w\left(t,y,x,z\right)=-\exp\left(-\rho\left(z+\theta\left(t,x,y\right)\right)\right)\,.
\end{align}
Then for all $(t,x,y,z) \in [0,T] \times \mathbb R^d \times \mathbb R^r \times \mathbb R\ $ and $\ u = (u_s)_{s \in [t,T]} \in \mathcal A_t$, we have 
\begin{align}
\label{eq:subopt}
\mathbb{E}\left[-\exp\left(-\rho\left(z_{T}^{t,x,z,u}-\left(\begin{matrix}
x_{T}^{t,x}\\
y_{T}^{t,y,u}
\end{matrix}\right)^{\intercal}\left(\begin{matrix}
\varPsi & \frac{1}{2}\varUpsilon\\
\frac{1}{2}\varUpsilon^{\intercal} & \Gamma
\end{matrix}\right)\left(\begin{matrix}
x_{T}^{t,x}\\
y_{T}^{t,y,u}
\end{matrix}\right)\right)\right)\right] \le w(t,x,y,z)\,.
\end{align}

Let Assumption \ref{assume:1} hold. If Assumption \ref{assume:2} also holds,  then equality is obtained in \eqref{eq:subopt} with the optimal control $(u^\star_s)_{s \in [t,T]} \in \mathcal A_t$ in feedback form 
\begin{align}
\label{eq:optcontrolsec1}
u_{s}^{\star}=\frac{1}{2}\,A\left(s\right)^{-1}\left(M_{2}\left(s\right)^{\intercal}x_{s}+2\,M_{3}\left(s\right)y_{s}-B\left(s\right)x_{s}\right)\,.
\end{align}
Otherwise, if Assumption \ref{assume:2} does not hold, then $\exists \, \overline \rho > 0$ such that $\forall \rho \in (0, \overline\rho],$ equality is obtained in \eqref{eq:subopt} with the optimal control \eqref{eq:optcontrolsec1}. Finally, in both cases, $w=\vartheta$ on $[0,T] \times \mathbb R \times \mathbb R^d \times \mathbb R^r.$
\end{thm}

Theorem \ref{thm1} solves the control problem \eqref{eq:vfdef} when the DRE \eqref{eq:RiccatiGeneral} with terminal condition \eqref{eq:termcondP} admits a unique solution on $[0, T]$. The optimal control is a linear feedback with gains obtained from $P$. Subsection \ref{sec:subsec:existence} uses the equivalent control problem to find a-priori bounds and proves existence and uniqueness of a solution to the DRE \eqref{eq:RiccatiGeneral}. 

\subsection{Existence and uniqueness of a solution \label{sec:subsec:existence}}

The next theorem, for which a proof is given in \ref{appendixproofthm2}, provides conditions for the existence and uniqueness of a solution to the DRE \eqref{eq:RiccatiGeneral} with terminal condition \eqref{eq:termcondP}.

\begin{thm}
\label{thm2}
Let Assumption \ref{assume:1} hold. If Assumption \ref{assume:2} also holds,  then  the DRE \eqref{eq:RiccatiGeneral} with terminal condition \eqref{eq:termcondP} has a unique symmetric solution. Otherwise, if Assumption \ref{assume:2} does not hold, then $\exists \, \overline \rho > 0$ such that  the DRE \eqref{eq:RiccatiGeneral} with terminal condition \eqref{eq:termcondP} has a unique symmetric solution for all $\rho \in (0, \overline\rho]$.

\end{thm}

The following results extend Theorem \ref{thm2} and show that solutions to the DRE \eqref{eq:RiccatiGeneral} provide solutions to larger families of DREs.
\begin{crl}\label{rem:1}
Assume $P$ is the unique solution to \eqref{eq:RiccatiGeneral}, then  the DRE 
\begin{align*}
\dot{\tilde{P}}\left(t\right) & =\tilde{Q}\left(t\right)+\tilde{Y}\left(t\right)^{\intercal}\tilde{P}\left(t\right)+\tilde{P}\left(t\right)\tilde{Y}\left(t\right)+\tilde{P}\left(t\right)U\left(t\right)\tilde{P}\left(t\right)\,
\end{align*} with terminal condition $\,\tilde P\left(T\right) = P\left(T\right)-K$ admits a unique solution, where $K \in \mathcal M_{d+r}\left(\mathbb R\right),$ and the coefficients are $\tilde{Y}\left(t\right)= Y\left(t\right)+U\left(t\right)K\,$, and $\ \tilde{Q}\left(t\right)= Q\left(t\right)+K\,U\left(t\right)K +Y\left(t\right)^{\intercal}K+K\,Y\left(t\right)\,$.
\end{crl}

\begin{crl}\label{rem:2}
Assume $P$ is the unique solution to \eqref{eq:RiccatiGeneral} and let $Z\in \mathcal C^1\left([0,T], \mathcal S_{d+r}^{++}\left(\mathbb R\right)\right)$, then  the DRE 
\begin{align*}
\dot{\utilde{P}}\left(t\right) & =\utilde{Q}\left(t\right)+\utilde{Y}\left(t\right)^{\intercal}\utilde{P}\left(t\right)+\utilde{P}\left(t\right)\utilde{Y}\left(t\right)+\utilde{P}\left(t\right)\utilde{U}\left(t\right)\utilde{P}\left(t\right)\,
\end{align*} with terminal condition  $\ \ \utilde P\left(T\right)=Z\left(T\right)^{\intercal}P\left(T\right)Z\left(T\right)\ \ $ admits a unique solution, where 
\begin{align*}
\begin{cases}
\utilde{Y}\left(t\right)= & Z\left(t\right)^{-1}Y\left(t\right)Z\left(t\right)+Z\left(t\right)^{-1}Z'\left(t\right)\ ,\\
\utilde{U}\left(t\right)= & Z\left(t\right)^{-1}U\left(t\right)\left(Z\left(t\right)^{\intercal}\right)^{-1}\ ,\\
\utilde{Q}\left(t\right)= & Z\left(t\right)^{\intercal}Q\left(t\right)Z\left(t\right)\, .
\end{cases}
\end{align*}

\end{crl}

In the next sections, we study two algorithmic portfolio trading problems which are solved using the result in Theorem \ref{thm2}. First, we extend multi-asset market making to dynamics that incorporate signals and hedging instruments. Next, we solve optimal portfolio trading with Bayesian learning of the drift.

\section{Multi-asset optimal market making \label{sec:begvou}}

We consider a market maker in charge of quoting multiple assets in an OTC market. She receives RFQs from different clients with variable sizes, and decides on the optimal quotes at any point in time. The market maker trades assets that she quotes and other hedging instruments in a dealer-to-dealer venue to hedge risk. The joint dynamics of the asset midprices, the hedging instruments prices, and a set of signals follow a multi-OU process. We use the approximation method introduced in \cite{bergault2018closed} to obtain a quoting and hedging strategy that can efficiently be computed by solving the DRE \eqref{eq:RiccatiGeneral}.

\subsection{The model}

Let $\left(\Omega, \mathcal F, \mathbb P; \mathbb F = (\mathcal F(t))_{t \in [0,T]} \right)$ be a standard filtered probability space that supports all the processes we introduce and satisfies the usual conditions. A market maker operates in an OTC market and is in charge of a portfolio containing $r \in \mathbb N^\star$ assets. She uses a set of $k \in \mathbb N$ signals and we model the joint dynamics of prices and signals by a $d$-dimensional multi-OU process $(S_t)_{t\in [0,T]} = \left(S^1_t, \ldots, S^d_t \right)^\intercal_{t\in [0,T]}$ with dynamics:
\begin{align} \label{PriceProces}
    dS_t &  = R\,(\overline S - S_t) \,dt + V\, dW_t\,, 
\end{align}
where $d= r + k$, the initial values $S_0 \in {\mathbb R}^d $ are known, $\overline S \in {\mathbb R}^d $, $R \in \mathcal M_d (\mathbb R)$, $V \in \mathcal M_{d,j}(\mathbb R)$, and $(W_t)_{t \in [0,T]} = \left(W^1_t, \ldots, W^j_t \right)^\intercal_{t \in [0,T]}$ is a $j$-dimensional standard Brownian motion with independent coordinates for some $j \in \mathbb{N}^\star$. Without loss of generality, we assume that the first $r$ elements of the vector $S$ correspond to the asset prices.

Multi-OU dynamics are effective when the market maker uses the predictive power of signals, or for hedging purposes when there exists a linear combination of assets that reduces the overall risk. In the dynamics \eqref{PriceProces}, the parameter $\overline{S}$ is the unconditional expectation of prices and signals in the long-run, and the matrix $R$ drives the joint deterministic part in the dynamics. We denote by $\Sigma = V\,V^\intercal$ the variance-covariance matrix associated with the process $(S_t)_{t\in [0,T]}$, and $\Sigma^r$ the variance-covariance matrix associated with the first $r$ elements of the process $(S_t)_{t\in [0,T]}$.

In practice, market makers hedge their positions with securities that they quote and with other liquid instruments for which they do not offer quotes, but which offset their risk. For example, a market maker in charge of corporate bonds hedges rates risk with bond futures, and further hedges credit risk with credit derivatives. Moreover, market makers  use latent factor models and other market variables with predictive power to enhance performance. Our framework is designed for these cases. 

To incorporate hedging instruments  in our model, for which the market maker does not propose quotes, it is enough to consider zero RFQ arrival intensities for these assets.  To incorporate signals, we consider joint dynamics of prices and signals, but only consider inventory in the assets; see \cite{cartea2019trading} for a similar setup.





\subsection{The dynamics}
Throughout the trading window $[0, T]$, the market maker interacts with $N$ clients and chooses the prices at which she is ready to buy or sell the quoted assets.

\paragraph{\textbf{Bid and ask quotes}} The market maker quotes depend on the client and  the size $z>0$ of the request. We define the bid and ask prices of asset $i$ for client's tier $n$, as the $\mathcal{P\otimes\mathcal{B}}\left(\mathbb{R}_{+}^{*}\right)$-measurable maps $S^{i,n,b},S^{i,n,a}:\Omega\times[0,T]\times\mathbb{R}_{+}^{*} \mapsto \mathbb R$, where $\mathcal{P}$ is the $\sigma$-algebra of predictable subsets of $\Omega\times[0,T]$, and $\mathcal{B}\left(\mathbb{R}_{+}^{*}\right)$ are the Borelian sets of $\mathbb{R}_{+}^{*}$. 

For $i \in \{1,\dots,r\}$ and $n\in\{1,\dots,N\},$ the maps $(\delta_{t}^{i,n,b}(\cdot) )_{t \in [0,T]}$ and $(\delta_{t}^{i,n,a}(\cdot) )_{t \in [0,T]}$ are the shifts from the reference price $S^i$ for, respectively, the bid and the ask quotes and we write
\begin{align}
    \delta_{t}^{i,n,b}(z) = S_{t}^{i} - S_{t}^{i,n,b}(z) \ \ \ \ \textrm{and} \ \ \ \  \delta_{t}^{i,n,a} = S_{t}^{i, n,a}(z) - S_{t}^{i} \ , \ \ \forall t \in [0,T].
\end{align}

\paragraph{\textbf{Number of transactions and RFQ sizes}} We define the probability measures $\psi^{i,n,b}\left(dz\right)$ and $\psi^{i,n,a}\left(dz\right)$ that correspond to the distributions of the requested sizes at the bid and ask side, respectively, of client's tier $n$ for asset $i$, and which are used to integrate over the distribution of sizes. 

For $i \in \{1,\dots,r\}$ and $n \in \{1,\dots,N\},$ we use two càdlàg $\mathbb{R_{+}^{*}}$-marked point processes $N^{i,n,b}\left(dt, dz\right)$ and $N^{i,n,a}\left(dt, dz\right)$ to model the number of transactions at the bid and ask side, for asset $i$ and tier $n$.  To construct the marked point processes, consider a new filtered probability space $\left(\Omega, \mathcal F, \tilde{\mathbb P}, \mathbb F \right)$ and two independent compound Poisson processes $M^{i,n,b}$ and $M^{i,n,a}$ with intensity one and with increments that follow the size distributions  $\psi^{i,n,b}\left(dz\right)$ and $\psi^{i,n,a}\left(dz\right)$. Let $N^{i,n,b}\left(dt, dz\right)$ and $N^{i,n,a}\left(dt, dz\right)$ be the associated random measures. For each $(\delta_{t}^{i,n,b}(\cdot) )_{t \in [0,T]}$ and $(\delta_{t}^{i,n,a}(\cdot) )_{t \in [0,T]}$, let $\mathbb P^\delta$ be a probability measure given by the Radon-Nikodym derivative $$\frac{\text{d}\mathbb{P}^{\delta}}{\text{d}\tilde{\mathbb{P}}}\Bigg|_{\mathcal{F}_{t}}=\mathcal{E}\left(\int_{\mathbb{R}_{+}^{\star}}\left(\varLambda^{i,n,b}\left(\delta_{t}^{i,n,b}(z)\right)-1\right)\tilde N^{i,n,b}\left(dt,dz\right)+\int_{\mathbb{R}_{+}^{\star}}\left(\varLambda^{i,n,a}\left(\delta_{t}^{i,n,a}(z)\right)-1\right)\tilde N^{i,n,a}\left(dt,dz\right)\right),$$ where $\mathcal E$ is the stochastic exponential and $\tilde N^{i,n,b}$ and $\tilde N^{i,n,a}$ are the compensated processes associated with $N^{i,n,b}$ and $N^{i,n,b}$, respectively. Thus, \cite{bremaud1977processus} show that under $\mathbb{P}^{\delta}$, the marked point processes $N^{i,n,b}\left(dt, dz\right)$ and $N^{i,n,a}\left(dt, dz\right)$ have intensity kernels $(\lambda_{t}^{i,n,b}\left(dz\right))_{t\in\mathbb{R_{+}}}$ and $\left(\lambda_{t}^{i,n,a}\left(dz\right)\right)_{t\in\mathbb{R_{+}}}$, respectively, given by
\begin{equation*}
    \lambda_{t}^{i,n,b}(dz)=\varLambda^{i,n,b}(\delta_{t}^{i,n,b}(z))\,\psi^{i,n,b}(dz)\,\ \ \ \, \text{and} \ \ \ \, \lambda_{t}^{i,n,a}(dz) =\varLambda^{i,n,a}\left(\delta_{t}^{i,n,a}(z)\right)\psi^{i,n,a}(dz)\,.
\end{equation*}
Recall that the intensities corresponding to the hedging instruments are set to zero. 

Moreover, for each asset $i \in \{1,\dots,r\}$ and tier $n \in \{1,\dots,N\}$,  we assume that the functions $\varLambda^{i,n,.}$ for the bid and the ask satisfy the following properties, where ${\varLambda^{i,n,.}}^{'}$ and ${\varLambda^{i,n,.}}^{''}$ are respectively the first and second derivative of $\varLambda^{i,n,.}$: (i) $\varLambda^{i,n,.}$ is twice continuously differentiable, (ii) ${\varLambda^{i,n,.}}^{'} (\delta) < 0$, (iii) $\lim\limits_{\delta \rightarrow +\infty} \varLambda^{i,.} (\delta) = 0$, and (iv) $\ \underset{\delta}{\sup} \frac{\varLambda^{i,n,.}(\delta) {\varLambda^{i,n,.}}^{''}(\delta)}{\left({\varLambda^{i,n,.}}^{'}(\delta)\right)^2 } <2 $.


\paragraph{\textbf{Inventory and internalisation / externalisation}} For $i \in \{1,\dots,r\}$, the inventory process of the market maker for asset $i$ is denoted by $(q_{t}^{i})_{t \in [0, T]}.$ We denote by $(q_{t})_{t \in [0, T]}$ the vector process $(q_t^{1},\ldots,q_t^{r})_{t\in [0, T]}^\intercal$. The market maker uses another trading venue to trade the assets for which she proposes quotes, but also to trade other hedging instruments. After an RFQ is received and the trade is executed, the market maker can either internalise or externalise. To internalise, the market maker keeps the trade in her inventory (while waiting for an offsetting order). To externalise, the market maker hedges risk by instantaneously offsetting the trade fully or partly, at a cost, in a dealer-to-dealer trading venue. We use the process $\left(v_{t}\right)_{t\in\mathbb{R_{+}}}=\left(v_{t}^{1},\dots,v_{t}^{r}\right)_{t\in\mathbb{R_{+}}}^\intercal$ to model the trading speed of the market maker in the dealer-to-dealer market, and we consider that the execution costs from her trading activity are quadratic in the speed. Finally, the inventory of the market maker in asset $i \in \{1,\dots,r\}$ has dynamics combining the results of her market making and hedging activity: 
\begin{align*}
    dq_{t}^{i}&=v_{t}^i\,dt + \sum_{n=1}^{N}\int_{\mathbb{R_{+}^{*}}}z N^{i,n,b}\left(dt,dz\right)-\sum_{n=1}^{N}\int_{\mathbb{R_{+}^{*}}}z N^{i,n,a}\left(dt,dz\right),\ \ q_{0}^{i}\ \ \textrm{given}.
\end{align*}

\paragraph{\textbf{Cash process}} We use the process $(X_{t})_{t \in [0, T]}$ to model the amount on the market maker's cash account and which follows the dynamics 
\begin{align*}
    dX_{t}=&-S_{t}^{\intercal} \mathcal J^\intercal dq_{t}- v_{t}^\intercal \eta v_t dt + \sum_{i=1}^{r}\sum_{n=1}^{N}\int_{\mathbb{R_{+}^{*}}}z\left(\delta_{t}^{i,n,b}\left(z\right)N^{i,n,b}\left(dt,dz\right)+\delta_{t}^{i,n,a}\left(z\right)N^{i,n,a}\left(dt,dz\right)\right),
\end{align*}
with $X_0 \in \mathbb R$ given, $\eta \in \mathcal S_r^{++} \left(\mathbb R\right)$ quantifies the execution costs from externalisation, and $\mathcal J$ is an $r \times d$ matrix with $\mathcal J_{ij} = \mathbbm 1_{i=j}\,.$ The matrix $\mathcal J$ maps the first $r$ elements of the $d$-dimensional price $S$ to an $r$-dimensional vector.





\subsection{The optimisation problem}

The market maker maximises the expected exponential utility of her final wealth by the end of the trading window $[0,T]$: 
\begin{align}\label{eq:mmperfcrit}
    \mathbb{E}\left[-\exp\left(-\rho\left(X_{T}+q_{T}^{\intercal}\,\mathcal J \,S_{T}-q_{T}^{\intercal}\,\Gamma \,q_{T}\right)\right)\right],
\end{align}
where $\mathcal A = \mathcal A_0$ is the set of admissible strategies, and $\mathcal A_t$ is defined as
\begin{align*}
    \mathcal{A}_t=\bigg\{\left(\delta^{b}_{s},\delta^{a}_{s},v_{s}\right)_{s \in [t,T]}:\Omega\times[t,T]&\times\mathbb{R}_{+}^{*}\rightarrow\mathbb{R^{\textrm{2}}\times\mathbb{R}^{r}}\quad\textrm{s.t.}\quad\left(\delta^{b},\delta^{a}\right)\textrm{ is }\mathcal{P}\otimes\mathcal{B}\left(\mathbb{R}_{+}^{*}\right)\textrm{-measurable} \\& \textrm{ and bounded from below, and } v \textrm{ is }\mathcal{P}\textrm{-measurable and bounded}\bigg\}.
\end{align*}

The final wealth in the performance criterion \eqref{eq:mmperfcrit} is the sum of the final cash amount $X_T$ and the remaining inventory valued at $q_T^{\intercal} \mathcal J S_T - q_T^\intercal \, \Gamma \, q_T$, where $\Gamma \in \mathcal S_r^{++} (\mathbb R)$. The final penalty $q_T^\intercal\, \Gamma \, q_T$ is a discount applied to the terminal mark-to-market value of the assets which penalises any non-zero final inventory. Note that in the case of hedging instruments with zero intensity, i.e., for which the market maker does not propose quotes, the performance \eqref{eq:mmperfcrit} of the market maker does not depend on the quoting strategy. In particular, any admissible control for the hedging instruments yields the same performance.

The value function $\vartheta :[0,T]\times\mathbb R \times \mathbb R^r \times \mathbb R^d \rightarrow \mathbb R$ of the market maker is 
\begin{align*}
    \vartheta(t,x,q,S) = \sup_{\substack{ \left(\delta^{b}, \delta^{a}, v\right)  \in \mathcal A_t }} \mathbb E \left[ -e^{-\rho \left( X^{t,x,S,\delta^{b}, \delta^{a},v}_T + \left(q^{t,q,\delta^{b},\delta^{a},v}_T\right)^{\intercal} \mathcal J S^{t, S}_T - \left(q^{t,q,\delta^{b},\delta^{a},v}_T\right)^\intercal \Gamma q^{t,q,\delta^{b},\delta^{a},v}_T \right)}   \right], 
\end{align*} where the controlled processes $\left(q^{t,q,\delta^{b},\delta^{a},v}_T\right)_{s\geq t}$, $\ \left(X^{t,x,S,\delta^{b}, \delta^{a},v}_T\right)_{s\geq t}$, and $\left(S^{t,S}_T\right)_{s\geq t}$ start at $t$ with values $\left(q, x,S\right)\,$ for the state variables. The Hamilton--Jacobi--Bellman (HJB) equation associated with the market making problem is\footnote{The symbols for the value function $\vartheta$ and for the solution to the HJB $w$ are different unless one proves equality with a verification argument; see \cite{barzykin2023algorithmic} where the authors solve a similar problem. Below, we focus on an approximation problem for the optimal quotes.}

\begin{align}
\label{hjbuACOMPREHENSIVE}
    0=&\,\partial_{t}w\left(t,x,q,S\right)+\left(\overline{S}-S\right)^{\intercal}R^{\intercal}\nabla_{S}w\left(t,x,q,S\right)+\frac{1}{2}\textrm{Tr}\left(\Sigma D_{SS}^{2}w(t,x,q,S)\right)\nonumber\\+&\sup_{v}\left\{ v^{\intercal}\nabla_{q}w(t,x,q,S)-\left(v^{\intercal}\mathcal J S+v^\intercal \eta v \right)\partial_{x}w(t,x,q,S)\right\} \nonumber\\+&\sum_{i=1}^{r}\sum_{n=1}^{N}\int_{\mathbb{R_{+}^{*}}}\sup_{\delta^{i,n,b}}\varLambda^{i,n,b}(\delta_{t}^{i,n,b}(z))\left(w(t,x-z(S^{i}-\delta_{t}^{i,n,b}(z)),q+ze^{i},S)\nonumber-w\left(t,x,q,S\right)\right)\psi^{i,n,b}(dz)\nonumber\\+&\sum_{i=1}^{r}\sum_{n=1}^{N}\int_{\mathbb{R_{+}^{*}}}\sup_{\delta^{i,n,a}}\varLambda^{i,n,a}(\delta_{t}^{i,n,a}(z))\left(w\left(t,x+z (S^{i}+\delta_{t}^{i,n,a}(z)),q-ze^{i},S\right) -w\left(t,x,q,S\right)\right)\psi^{i,n,a}(dz),
\end{align}
where $\{e^i\}$ is the canonical basis of $\mathbb R^r$, with terminal condition 
\begin{align}
\label{TChjbACOMPREHENSIVE}
    w(T,x,q,S)=-\exp\left(-\rho\left(x+q^{\intercal}\mathcal J S-q^{\intercal}\Gamma q\right)\right), \quad \forall (x,q,S) \in \mathbb R \times \mathbb R^r \times \mathbb R^d.
\end{align}

In the next proposition, for which a proof is straightforward, we use the ansatz 
\begin{align*}
w\left(t,x,q,S\right)=-\exp\left(-\rho\left(x+q^{\intercal} \mathcal J  S+\theta\left(t,q,S\right)\right)\right)\,.
\end{align*}

\begin{prop}
Assume there is a solution $\theta \in \mathcal C^{1,1,2}\left([0, T] \times \mathbb R^r \times \mathbb R^d, \mathbb R\right)$ to the equation:
\begin{align}
\label{HJBthetaACOMPREHENSIVE}
    0=&\ \partial_{t}\theta\left(t,q,S\right)+\left(\overline{S}-S\right)^{\intercal}R^{\intercal}\left(\mathcal J^\intercal q+\nabla_{S}\theta\left(t,q,S\right)\right)+\frac{1}{2}\textrm{Tr}\left(\Sigma D_{SS}^{2}\theta\left(t,q,S\right)\right) \nonumber 
    \\
    &-\frac{\rho}{2}\left(\mathcal J^\intercal  q+\nabla_{S}\theta\left(t,q,S\right)\right)^{\intercal}\Sigma\left(\mathcal J^\intercal q+\nabla_{S}\theta\left(t,q,S\right)\right)+\sup_{v}\left\{ v^{\intercal}\nabla_{q}\theta\left(t,q,S\right)-v^\intercal \eta v \right\}  \nonumber
    \\
    &+\sum_{i=1}^{r}\sum_{n=1}^{N}\int_{\mathbb{R_{+}^{*}}}z H^{i,n,b}\left(z,\frac{\theta\left(t,q,S\right)-\theta\left(t,q+z e^{i},S\right)}{z}\right)\psi^{i,n,b}\left(dz\right)\nonumber\\
    &+\sum_{i=1}^{r}\sum_{n=1}^{N}\int_{\mathbb{R_{+}^{*}}}z H^{i,n,a}\left(z,\frac{\theta\left(t,q,S\right)-\theta\left(t,q-z e^{i},S\right)}{z}\right)\psi^{i,n,a}\left(dz\right),
\end{align}
on $[0, T) \times \mathbb R^r \times \mathbb R^d$, with terminal condition 
\begin{align} 
\label{TCthetaACOMPREHENSIVE}
    \theta(T, q, S) =  - q^\intercal \Gamma q, \quad \forall (q, S) \in \mathbb R^r \times \mathbb R^d,
\end{align}
where we define the Hamiltonian functions $H^{i,n,b}$ and $H^{i,n,a}$ respectively as 
\begin{align}
\label{defHamiltoniansCOMPREHENSIVE}
    H^{i,n,b}\left(z,p\right)=&\sup_{\delta}\frac{\varLambda^{i,n,b}\left(\delta\right)}{\rho z}\left(1-\exp\left(-\rho z(\delta-p)\right)\right)\,, \nonumber \\
    H^{i,n,a}\left(z,p\right)=&\sup_{\delta}\frac{\varLambda^{i,n,a}\left(\delta\right)}{\rho z}\left(1-\exp\left(-\rho z(\delta-p)\right)\right)\,.
\end{align}
Then, the function $w:[0, T] \times \mathbb R \times \mathbb R^r \times \mathbb R^d \rightarrow \mathbb R$, defined by
\begin{align}
    \label{ansatzUCOMPREHENSIVE}
    w(t,x,q,S) = -\exp \left( -\rho \left( x + q^{\intercal} \mathcal J S + \theta (t,q,S) \right) \right), \ \ \forall (t,x,q,S) \in [0, T] \times \mathbb R \times \mathbb R^r \times \mathbb R^d\,,
\end{align}
is a solution to \eqref{hjbuACOMPREHENSIVE} with terminal condition \eqref{TChjbACOMPREHENSIVE}.
\end{prop}

To obtain the optimal trading (externalisation) speed in feedback form, we write the Legendre-Fenchel transform of the quadratic execution cost $v^\intercal\,\eta\,v$ as
\begin{align*}
    L: p \in \mathbb R^d \mapsto \sup_{v \in \mathbb R^d} v^\intercal p - v^\intercal \,\eta \,v  = \frac{1}{4}\, p^\intercal \,\eta^{-1}\, p,
\end{align*}
so the supremum of $ v^{\intercal}\nabla_{q}\theta\left(t,q,S\right)-v^\intercal \eta v $ in \eqref{hjbuACOMPREHENSIVE}  is reached at 
$$v_s^\star = \frac12 \nabla_q \theta\left(s,q_s,S_s\right)^\intercal \eta^{-1}.$$ Second, to obtain the optimal quotes in feedback form, similar arguments as in \cite{gueant2017optimal} lead to the optimal bid and ask quotes $S^{i,n,b}_t\left(z\right) = S^{i}_t - \delta^{i,n,b*}_t\left(z\right)$ and $S^{i,n,a}_t\left(z\right) = S^{i}_t + \delta^{i,n,a*}_t\left(z\right)$ for an RFQ of size $z$, where 
\begin{align}
\label{closedloopoptimalcontrols}
    \delta^{i,n,b*}_t\left(z\right) = \ & \tilde \delta^{i,n,b*} \left( z,  \frac{\theta(t,q_{t-},S_t) - \theta(t,q_{t-} +z e^i,S_t) }{z} \right) \, , \nonumber \\
    \delta^{i,n,a*}_t\left(z\right) = \ & \tilde \delta^{i,n,a*} \left( z,  \frac{\theta(t,q_{t-},S_t) - \theta(t,q_{t-} -z e^i,S_t) }{z} \right) \, ,
\end{align}
and the functions $\tilde \delta^{i,n,b*}(\cdot,\cdot)$ and $\tilde \delta^{i,n,a*}(\cdot,\cdot)$ are 
\begin{align}\label{eq:quotefunctions}
    {\tilde \delta^{i,n,b*}} (z, p) & = \varLambda^{i,n,b^{-1}} \left( \rho z H^{i,n,b}(z,p) - {H^{i,n,b}}'(z,p) \right),  \nonumber \\
    {\tilde \delta^{i,n,a*}} (z, p) & = \varLambda^{i,n,a^{-1}} \left( \rho z H^{i,n,a}(z,p) - {H^{i,n,a}}'(z,p) \right),
\end{align}
and ${H^{i,n,.}}'(z,p) $ denotes the derivative of $H^{i,n,.}(z,p) $ with respect to $p$. It is beyond the scope of this paper to provide a formal solution to the optimal quote problem and we focus on the approximation technique \citep[see][for a rigorous solution to a similar problem]{barzykin2023algorithmic}.

\subsection{Quadratic approximation of the value function \label{sec:approx}}

One usually employs numerical methods to approximate the value function and hence the optimal quotes \eqref{closedloopoptimalcontrols}.  These methods suffer from the curse of dimensionality and do not usually scale up to the multi-asset case, because the number of dimensions grows exponentially with the number of assets. In this subsection, we approximate the solution $\theta$ of the HJB \eqref{HJBthetaACOMPREHENSIVE} characterising the optimal quotes by the solution $\tilde \theta$ of another HJB for which computations are easier to carry out.

For small RFQ sizes, we use the approximation   $$\frac{\theta(t,q,S) - \theta(t,q+z e^i,S)}{z} \approx - \frac{\theta(t,q,S) - \theta(t,q-z e^i,S)}{z} \,.$$ Thus, the term  we wish to approximate in \eqref{HJBthetaACOMPREHENSIVE} becomes a function of the form $p \rightarrow H^{i,n,b}(p) + H^{i,n,a}(-p).$ Moreover, the functions $H^{i,n,b}$ and $H^{i,n,a}$ in \eqref{defHamiltoniansCOMPREHENSIVE} are positive and decreasing in $p$, so we use a quadratic approximation. Furthermore, a quadratic approximation simplifies the problem to finding the solution to a DRE of the form \eqref{eq:RiccatiGeneral}.  More precisely, for all $i \in \{1,\dots,r\}$ and $n \in \{1,\dots,N\}$, we approximate the Hamiltonian term 
\begin{align*}
    H^{i,n,b} \left(z, \frac{\theta(t,q,S) - \theta(t,q+z e^i,S) }{z} \right) + H^{i,n,a} \left(z, \frac{\theta(t,q,S) - \theta(t,q-z e^i,S) }{z} \right)
\end{align*}
by the sum of the two quadratic functions   
\begin{align*}
\begin{cases}
\tilde{H}^{i,n,a}:\left(z,p\right)\mapsto & \alpha_{0}^{i,n,a}\left(z\right)+\alpha_{1}^{i,n,a}\left(z\right)p+\frac{1}{2}\alpha_{2}^{i,n,a}\left(z\right)p^{2}\,,\\
\tilde{H}^{i,n,b}:\left(z,p\right)\mapsto & \alpha_{0}^{i,n,b}\left(z\right)+\alpha_{1}^{i,n,b}\left(z\right)p+\frac{1}{2}\alpha_{2}^{i,n,b}\left(z\right)p^{2}\ .
\end{cases}
\end{align*} 

The values of the approximation coefficients for each request size $z$ are obtained using the Taylor expansion around $p=0$: 
\begin{equation*}
\begin{cases}
\alpha_{0}^{i,n,b}(z) & ={H^{i,n,b}}(z,0)\,,\\
\alpha_{0}^{i,n,a}(z) & ={H^{i,n,a}}(z,0)\,,
\end{cases}\ \ \ \text{and}\ \ \ \begin{cases}
\alpha_{1}^{i,n,b}(z) & ={H^{i,n,b}}^{'}(z,0)\,,\\
\alpha_{1}^{i,n,a}(z) & ={H^{i,n,a}}^{'}(z,0)\,,
\end{cases}\ \ \ \text{and}\ \ \ \begin{cases}
\alpha_{2}^{i,n,b}(z) & ={H^{i,n,b}}^{''}(z,0)\,,\\
\alpha_{2}^{i,n,a}(z) & ={H^{i,n,a}}^{''}(z,0)\,.
\end{cases}
\end{equation*}

Next, we denote the approximation of the function $\theta$ by $\tilde \theta$, which verifies the HJB 
\begingroup
\allowdisplaybreaks
\begin{align}
\label{approxHJBCOMPRENHSIVE}
0=&\,\partial_{t}\tilde\theta\left(t,q,S\right)+\left(\overline{S}-S\right)^{\intercal}R^{\intercal}\left(\mathcal J^\intercal q+\nabla_{S}\tilde\theta\left(t,q,S\right)\right)+\frac{1}{2}\textrm{Tr}\left(\Sigma D_{SS}^{2}\tilde\theta\left(t,q,S\right)\right) \nonumber\\
&-\frac{\rho}{2}\left(\mathcal J^\intercal  q+\nabla_{S}\tilde\theta\left(t,q,S\right)\right)^{\intercal}\Sigma\left(\mathcal J^\intercal q+\nabla_{S}\tilde\theta\left(t,q,S\right)\right) +\frac{1}{4}\nabla_{q}\tilde\theta\left(s,q_{s},S_{s}\right)^{\intercal}\eta^{-1}\nabla_{q}\tilde\theta\left(s,q_{s},S_{s}\right) \nonumber\\
&+\sum_{i=1}^{r}\sum_{n=1}^{N}\int_{\mathbb{R_{+}^{*}}}z\alpha_{0}^{i,n,b}\left(z\right)\psi^{i,n,b}\left(dz\right)+\sum_{i=1}^{r}\sum_{n=1}^{N}\int_{\mathbb{R_{+}^{*}}}z\alpha_{0}^{i,n,a}\left(z\right)\psi^{i,n,a}\left(dz\right)\nonumber\\
&+\sum_{i=1}^{r}\sum_{n=1}^{N}\int_{\mathbb{R_{+}^{*}}}\alpha_{1}^{i,n,b}\left(z\right)\left(\tilde{\theta}\left(t,q,S\right)-\tilde{\theta}\left(t,q+z e^{i},S\right)\right)\psi^{i,n,b}\left(dz\right)\nonumber\\
&+\sum_{i=1}^{r}\sum_{n=1}^{N}\int_{\mathbb{R_{+}^{*}}}\alpha_{1}^{i,n,a}\left(z\right)\left(\tilde{\theta}\left(t,q,S\right)-\tilde{\theta}\left(t,q+z e^{i},S\right)\right)\psi^{i,n,a}\left(dz\right)\nonumber\\
&+\sum_{i=1}^{r}\sum_{n=1}^{N}\int_{\mathbb{R_{+}^{*}}}\frac{1}{2z}\alpha_{2}^{i,n,b}\left(z\right)\left(\tilde{\theta}\left(t,q,S\right)-\tilde{\theta}\left(t,q+z e^{i},S\right)\right)^{2}\psi^{i,n,b}\left(dz\right)\nonumber\\
&+\sum_{i=1}^{r}\sum_{n=1}^{N}\int_{\mathbb{R_{+}^{*}}}\frac{1}{2z}\alpha_{2}^{i,n,a}\left(z\right)\left(\tilde{\theta}\left(t,q,S\right)-\tilde{\theta}\left(t,q+z e^{i},S\right)\right)^{2}\psi^{i,n,a}\left(dz\right)\,,
\end{align}
\endgroup
with terminal condition 
\begin{align}
\label{approxTCCOMPREHNSIVE}
\tilde \theta(T, q, S) =  - q^\intercal \Gamma q, \quad \forall (q,S) \in \mathbb R^r \times \mathbb R^d.
\end{align}

To further simplify the above HJB, use the ansatz 
\begin{align}
\label{ansatzThetaTildeCOMPREHENSIVE}
    \tilde\theta(t,q,S)=q^{\intercal}A(t)q+q^{\intercal}B(t)S+S^{\intercal}C(t)S+D(t){}^{\intercal}q+E(t)^{\intercal}S+F(t),
\end{align} 
based on the next proposition whose proof is omitted.

\begin{prop}
{\allowdisplaybreaks
Assume that there exist matrix functions $A \in C^1 \left([0,T], \mathcal S_r(\mathbb R) \right)$, $B \in C^1 \left([0,T], \mathcal M_{r,d}(\mathbb R) \right)$, $C \in C^1 \left([0,T], \mathcal S_d(\mathbb R) \right)$, $D \in C^1 \left([0,T], \mathbb R^r \right)$, $E \in C^1 \left([0,T], \mathbb R^d \right)$, $F \in C^1 \left([0,T], \mathbb R \right)$ satisfying the system of ODEs
\begin{align}
\label{ODEACOMPREHENSIVE}
\begin{cases}
-\dot{A}(t)= & -\frac{\rho}{2}\,\left(B(t)-\mathcal{J}\right)\,\Sigma\,\left(B(t)^{\intercal}-\mathcal{J}\right)+A(t)\,\left(2\,V_{2,1}^{b}+2\,V_{2,1}^{a}+\eta^{-1}\right)\,A(t)\\
-\dot{B}(t)= & \left(\mathcal{J}-B(t)\right)\left(-R+2\,\rho\,\Sigma\,C(t)\right)+A(t)\left(2\,V_{2,1}^{b}+2\,V_{2,1}^{a}+\eta^{-1}\right)\,B(t)\\
-\dot{C}(t)= & R^{\intercal}C(t)+C(t)\,R-2\,\rho\,C(t)\,\Sigma\,C(t)+\frac{1}{4}B(t)^{\intercal}\left(2\,V_{2,1}^{b}+2\,V_{2,1}^{a}+\eta^{-1}\right)\,B(t)\\
-\dot{D}(t)= & \left(\mathcal{J}-B(t)\right)\left(R\,\overline{S}+\rho\,\Sigma\,E(t)\right)+2\,A(t)\,\left(v_{1,1}^{b}-v_{1,1}^{a}\right)\\
 & +2\,A(t)\,\left(V_{2,2}^{b}-V_{2,2}^{a}\right)\,\mathcal{D}\left(A(t)\right)+2\,A(t)\left(V_{2,1}^{b}+V_{2,1}^{a}\right)\,D(t)+A(t)^{\intercal}\,\eta^{-1}\,D(t)\\
-\dot{E}(t)= & -2\,C(t)\,R\,\overline{S}+R^{\intercal}\,E(t)-2\,\rho\,C(t)\,\Sigma\,E(t)+B(t)^{\intercal}\left(v_{1,1}^{b}-v_{1,1}^{a}\right)\\
 & +B(t)^{\intercal}\left(V_{2,2}^{b}-V_{2,2}^{a}\right)\mathcal{D}\left(A(t)\right)+B(t)^{\intercal}\left(V_{2,1}^{b}+V_{2,1}^{a}\right)D(t)+\frac{1}{2}\,B(t)^{\intercal}\,\eta^{-1}\,D(t)\\
-\dot{F}(t)= & -\overline{S}^{\intercal}R^{\intercal}E(t)-\frac{\rho}{2}\,E(t)^{\intercal}\,\Sigma\,E(t)-\textrm{Tr}\left(\Sigma\,C(t)\right)+\textrm{Tr}\left(V_{0,1}^{b}+V_{0,1}^{a}\right)+\textrm{Tr}\left(\left(V_{1,2}^{b}+V_{1,2}^{a}\right)A(t)\right)\\
 & +\left(v_{1,1}^{b}-v_{1,1}^{a}\right)^{\intercal}D(t)+\frac{1}{2}\,D(t)^{\intercal}\left(V_{2,1}^{b}+V_{2,1}^{a}\right)\,D(t)+\frac{1}{2}\,\mathcal{D}\left(A(t)\right)^{\intercal}\left(V_{2,3}^{b}+V_{2,3}^{a}\right)\,\mathcal{D}\left(A(t)\right)\\
 & +\mathcal{D}\left(A(t)\right)^{\intercal}\left(V_{2,2}^{b}-V_{2,2}^{a}\right)\,D(t)+\frac{1}{4}\,D(t)^{\intercal}\,\eta^{-1}\,D(t)\,,
\end{cases}
\end{align}}
with terminal conditions 
\begin{align}
\label{termcondABCDEFCOMPREHENSIVE}
A(T) = -\Gamma\,, \quad B(T) = 0\,, \quad C(T) = 0\,, \quad D(T) = 0\,, \quad E(T) = 0\,, \quad F(T)=0\,,
\end{align}
where $\mathcal D$ is the linear operator mapping a matrix onto the vector of its diagonal coefficients, and for $i\in \{1,\ldots,r\}$, $n \in \{1,\dots,N\}$, $j\in\{0,1,2\}$, and $m\in\mathbb{N}$:
\begin{equation*}
\begin{cases}
\Delta_{j,m}^{i,n,b}=\int_{\mathbb{R_{+}^{*}}}z^{m}\alpha_{j}^{i,n,b}(z)\psi^{i,n,b}(dz)\,,\\
\Delta_{j,m}^{i,n,a}=\int_{\mathbb{R_{+}^{*}}}z^{m}\alpha_{j}^{i,n,a}(z)\psi^{i,n,b}(dz)\,,
\end{cases}\text{and}\ \ \begin{cases}
v_{j,m}^{b}=\left(\sum\limits _{n=1}^{N}\Delta_{j,m}^{i,n,b}\right)_{i\in\{1,\dots,r\}}^{\intercal}\,,\\
v_{j,m}^{a}=\left(\sum\limits _{n=1}^{N}\Delta_{j,m}^{i,n,a}\right)_{i\in\{1,\dots,r\}}^{\intercal}\,,
\end{cases}\text{and}\ \ \begin{cases}
V_{j,m}^{b}=\textrm{diag}(v_{j,m}^{b})\,,\\
V_{j,m}^{b}=\textrm{diag}(v_{j,m}^{a})\,.
\end{cases}
\end{equation*}

Then, $\tilde \theta$ defined by \eqref{ansatzThetaTildeCOMPREHENSIVE} satisfies \eqref{approxHJBCOMPRENHSIVE} with terminal condition \eqref{approxTCCOMPREHNSIVE}.

\end{prop}

Notice that the  system of ODEs  \eqref{ODEACOMPREHENSIVE} can be solve sequentially. First, solve the subsystem in $\left(A, B, C\right)$, then solve the linear subsystem in $\left(D, E\right)$, and $F$ is obtained by an integration. Thus, obtaining the approximated $\tilde \theta$ reduces to solving the subsystem in $\left(A, B, C\right)$. Define $P : [0,T] \rightarrow \mathcal S_{d+r}(\mathbb R)$ as 
\begin{align}
\label{PmatdefinitionCOMPREHENSIVE}
P(t) = \begin{pmatrix} A(t) & \frac 12 \, B(t) \\ \frac 12 \, B(t)^\intercal & C(t) \end{pmatrix}\,,
\end{align}
and observe that the ODE system  in $\left(A, B, C\right)$ is equivalent to the DRE 
\begin{align}
\label{PODECOMPREHENSIVE}
\dot{P}(t) = Q + Y^\intercal P(t) +  P(t) \, Y +  P(t) \, U \, P(t) \,,
\end{align}
with terminal condition  \begin{align}\label{PODECOMPREHENSIVETC}
 P(T) = \begin{pmatrix} -\Gamma & 0 \\ 0 & 0 \end{pmatrix} \,,
\end{align}
where  
\begin{align*}
Q=\frac{1}{2}\begin{pmatrix}\rho\,\Sigma^r & \mathcal J \, R\\
R^{\intercal} \, \mathcal J^\intercal & 0
\end{pmatrix}\,,\quad Y=\begin{pmatrix}0 & 0\\
\rho\,\Sigma\, \mathcal J^\intercal & R
\end{pmatrix}\,,\quad\text{and}\quad U=\begin{pmatrix}-2\left(V_{2,1}^{b}+V_{2,1}^{a}\right)-\eta^{-1} & 0\\
0 & 2\,\rho\,\Sigma
\end{pmatrix}\,.
\end{align*}

Assume that $\ 2V_{2,1}^b + 2V_{2,1}^a + \eta^{-1} \in \mathcal S_r^+ \left(\mathbb R\right)\,$. Here, Assumptions \ref{assume:1}  and \ref{assume:2} hold, thus the results of Section \ref{sec:riccati} apply to the DRE \eqref{PODECOMPREHENSIVE} with terminal condition \eqref{PODECOMPREHENSIVETC}. In particular, use Corollary \ref{rem:1} with $K=-\frac{1}{2}\left(\begin{matrix}
0 & \mathcal J\\
\mathcal J^\intercal & 0
\end{matrix}\right)$ and $\varUpsilon = \mathcal J$ in the terminal condition \eqref{eq:termcondP} to obtain the DRE \eqref{PODECOMPREHENSIVE}.

The approximation method of this section leads to the DRE \eqref{PODECOMPREHENSIVE} for which a solution exists and is unique, and which can be approximated rapidly  using existent efficient solving techniques. Once the approximated value function is obtained, the approximated quotes are
\begin{align}
\label{closedloopoptimalcontrols_approx}
    \delta^{i,n,b*}_t\left(z\right) = \ & \tilde \delta^{i,n,b*} \left( z,  \frac{\theta(t,q_{t-},S_t) - \theta(t,q_{t-} +z e^i,S_t) }{z} \right) \, , \nonumber \\
    \delta^{i,n,a*}_t\left(z\right) = \ & \tilde \delta^{i,n,a*} \left( z,  \frac{\theta(t,q_{t-},S_t) - \theta(t,q_{t-} -z e^i,S_t) }{z} \right) \, ,
\end{align}
where the functions $\tilde \delta^{i,n,b*}$ and $\tilde \delta^{i,n,a*}$ are in \eqref{eq:quotefunctions}. Moreover, the externalisation strategy associated with the approximated value function is 
\begin{align}
\label{optcontrol}
\overline v^\star_t = \frac 12 \eta^{-1} \left( 2A(t) q_t + B(t) S_t + D(t)  \right)\,.
\end{align}

\paragraph{\textbf{The case of symmetric exponential intensity functions}} If the intensity functions for the bid and ask sides are exponential, which is a commonplace assumption in the market making literature, then for all $i \in \{1,\dots,r\}$ and $n \in \{1,\dots,N\}$, we write
\begin{align} \label{eq:formintensity}
    \varLambda^{i,n,b} \left(\delta\right) = \varLambda^{i,n,b} \left(\delta\right) = A^{i,n} e^{-k^{i,n} \delta},
\end{align}
where $A^{i,n} $ and $k^{i,n} $ are positive constants. Thus, the Hamiltonian functions are 
\begin{align*}
    H^{i,n,b}(z, p) = H^{i,n,a}(z, p) = \frac{A^{i, n}}{k^{i,n}} \left(1+ \frac{\rho z}{k^{i,n}} \right)^{-\left(1+\frac{k^{i,n}}{\rho z^ {i,n}}\right)} e^{-k^{i,n} p},
\end{align*}
and the approximated quotes are given by the shifts 
\begin{align}
    \label{ApproxDeltaExp2}
    \overline{\delta}^{i,n,b}\left(z\right)&=-2q_{t-}^{\intercal}A(t)e^{i}-z e^{i^{\intercal}}A(t)e^{i}-S^{\intercal}B(t)^{\intercal}e^{i}-e^{i^{\intercal}}D(t) + \frac{1}{\rho z} \log\left(1+\frac{\rho z}{k^{i,n}}\right)\,,\\ \nonumber
    \overline{\delta}^{i,n,a}\left(z\right)&=2q_{t-}^{\intercal}A(t)e^{i}-z e^{i^{\intercal}}A(t)e^{i}+S^{\intercal}B(t)^{\intercal}e^{i}+e^{i^{\intercal}}D(t) + \frac{1}{\rho z} \log\left(1+\frac{\rho z}{k^{i,n}}\right)\,.
\end{align}



\subsection{Numerical results}

\begin{figure}[H]
    \centering
    \includegraphics{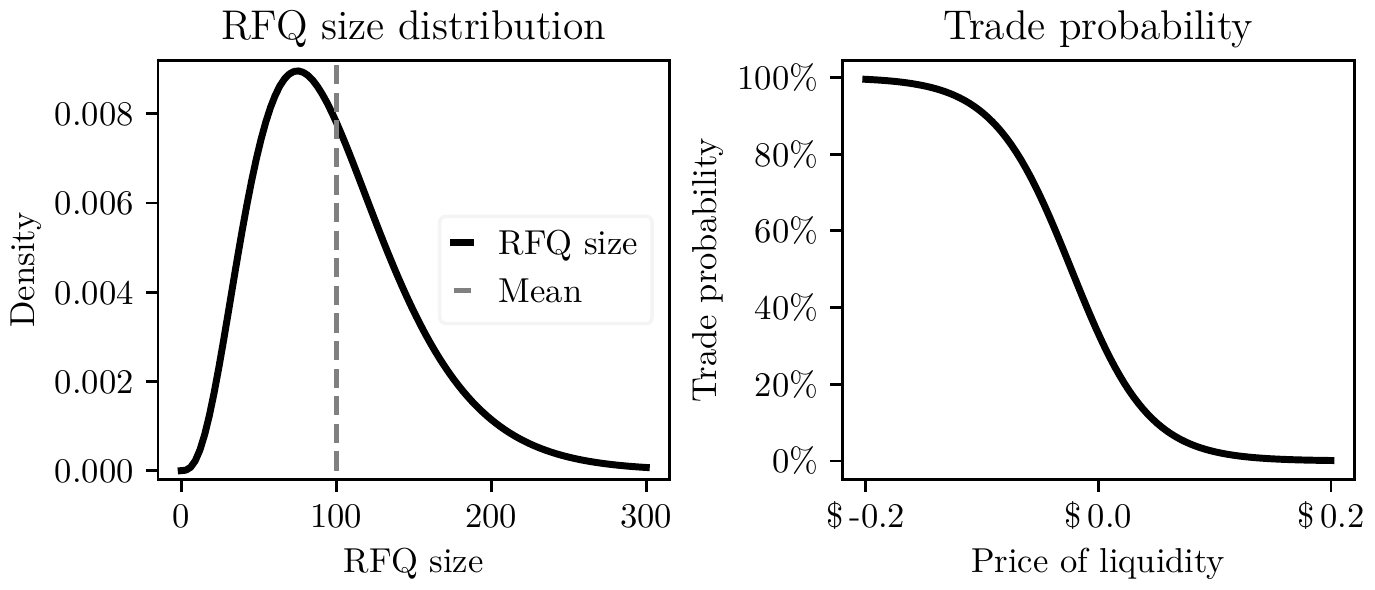}
    \caption{Left panel: distribution of RFQ sizes. Right panel:  probability of trading as a function of the price of liquidity.}
    \label{fig:prelim}
\end{figure}

We study the optimal quotes \eqref{closedloopoptimalcontrols} and the approximated quotes \eqref{closedloopoptimalcontrols_approx} for a market maker in charge of a single asset with mean reverting prices, and for a market maker in charge of two cointegrated assets. The market maker interacts with one client and we assume that a unique intensity function drives the trading flow at the bid and the ask and we write $$\Lambda^{b}\left(\delta\right) = \Lambda^{b}\left(\delta\right) = \lambda_{\text{base}}\,\frac{1}{1 + \exp\left(a+b\,\delta\right)}\,$$ for all the assets that we consider, where $\lambda_{\text{base}}$ is the base intensity, i.e., the intensity of order arrival when the price of liquidity is zero. Moreover, we assume that the RFQ sizes for all assets are distributed according to a Gamma distribution $\Gamma\left(\alpha, \beta\right)$. Table \ref{table:intensity} shows the parameter values for the intensity functions and the distribution of RFQ sizes, and Figure \ref{fig:prelim} shows the distribution of RFQ sizes and the probability to trade as a function of the price of liquidity implied by the parameter values of Table \ref{table:intensity}.

\begin{table}[H]
\begin{center}
\begin{tabular}{c |  c  c   c | c  c } 
Parameter & $\lambda_{\text{base}}$ & $a $ & $b$ & $\alpha$  & $\beta$  \\ [0.5ex] 
\hline
Value & $30\ \textrm{day}^{-1}$  & $0.7\ $ & $ 30\,\$^{-1}$ &  $ 4$ & $4 \cdot 10^{-2}$  \\ [0.5ex] 
\end{tabular}
\end{center}
\caption {Parameter values for the intensity function, and the RFQ size distribution.}
\label{table:intensity}
\end{table}

To compute the optimal and approximated bid and ask quotes, we discretise the Gamma distribution (left panel of Figure \ref{fig:prelim}) and we consider the sizes and the weights in Table \ref{table:gammadiscrete}. Finally, we assume that the market maker only trades if her inventory is within the range $[-600, 600].$ 

\begin{table}[H]
\begin{center}
\begin{tabular}{c |  c  c   c  c  c   c  c   c  c  c } 
RFQ Size & $25$ & $50$ & $75$ & $100$ & $125$ & $150$ & $175$ & $200$ & $225$ & $250$  \\ [0.5ex] 
\hline
Weight & $6.2\,\%$ & $18.1\,\%$ & $22.5\,\%$ & $19.7\,\%$ & $14.1\,\%$ & $8.9\,\%$ & $5.2\,\%$ & $2.9\,\%$ & $1.6\,\%$ & $0.8\,\%$  \\ [0.5ex] 
\end{tabular}
\end{center}
\caption {Discretisation of the RFQ size distribution.}
\label{table:gammadiscrete}
\end{table}


\subsubsection{Single-asset case} 

We consider a market maker in charge of a single asset. Table  \ref{table:1DOUParams} shows the parameter values for the OU dynamics of the price $S$ in \eqref{PriceProces} and for the performance criterion in \eqref{eq:mmperfcrit}. In the single-asset case, the parameter $R$ drives the mean reversion speed of prices towards the long-term unconditional mean $\overline S.$ 

\begin{table}[H]
\begin{center}
\begin{tabular}{c | c  c  c  c  | c  c  c } 
Parameter & $S_0$ & $\overline S$ & $R$ &  $V$ & $T$ & $\Gamma$ & $\rho$ \\ [0.5ex] 
\hline
Value & $\$\,  100 $   & $\$ 100$ &  $0\,$ or $\,0.1\ \textrm{day}^{-1}$ &  $1.2\ \$ \cdot \textrm{day}^{- \frac 12}$ & $7\ \textrm{day}$ &  $\$\,0\ $ & $ 10^{-3}\ \textrm{\$} ^{-1}$  \\
\end{tabular}
\end{center}
\caption {Parameters values for the asset prices and for the performance criterion of the market maker.}
\label{table:1DOUParams}
\end{table}



To obtain the optimal shifts $\delta^{b,\star}$ and $\delta^{a,\star}$, we employ an implicit Euler scheme to approximate the solution $\theta$ of the HJB \eqref{HJBthetaACOMPREHENSIVE}.\footnote{The scheme is in dimension three; one  for time, one for inventory, and one for the asset price.} To obtain the approximated quotes \eqref{closedloopoptimalcontrols_approx}, we use an implicit Euler scheme to approximate the solution to the DRE \eqref{PODECOMPREHENSIVE}.\footnote{The code for the numerical examples is in \href{https://github.com/FDR0903/Market_Making_Riccati}{https://github.com/FDR0903/Market\_Making\_Riccati}.}
 
\begin{figure}[H]
    \centering
    \includegraphics{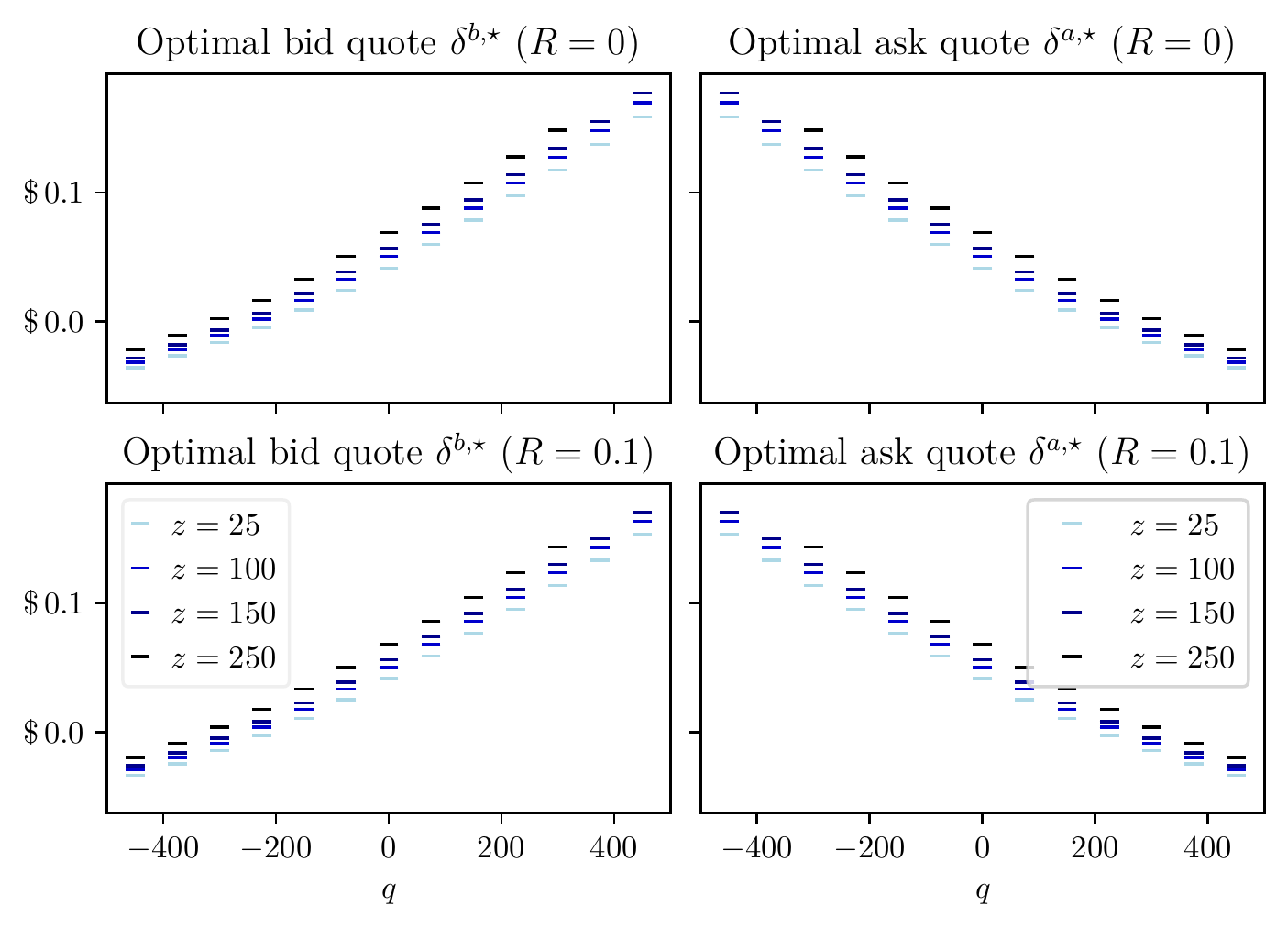}
    \caption{Optimal bid and ask shifts in \eqref{closedloopoptimalcontrols} as a function of the inventory and the RFQ size when $S=\overline S= 100$.}
    \label{fig:optimal_quotes_function_size}
\end{figure}

\paragraph{\textbf{Optimal quotes as a function of the RFQ size}} Figure \ref{fig:optimal_quotes_function_size} shows the optimal quotes for the different RFQ sizes in Table \ref{table:gammadiscrete}, and shows that the price of liquidity at the bid and the ask increases with the RFQ size.

\begin{figure}[H]
    \centering
    \includegraphics{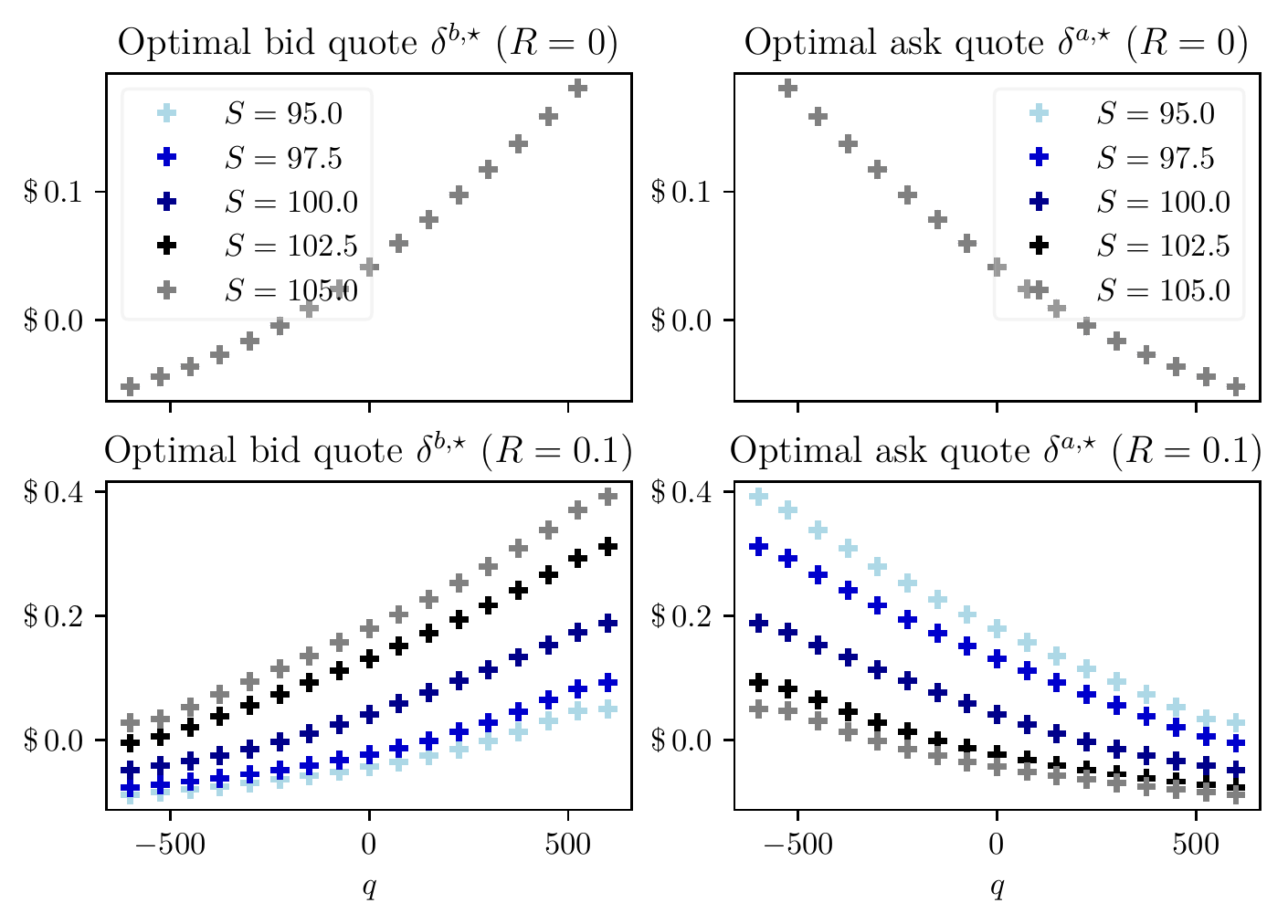}
    \caption{Optimal bid and ask quotes in \eqref{closedloopoptimalcontrols}  as a function of inventory $q$, price $S$, and mean reversion parameter $R$, when the RFQ size is $z=100$ shares.}
    \label{fig:optimal_quotes_function_S}
\end{figure}

\paragraph{\textbf{Optimal shifts as a function of the asset price}} Figure \ref{fig:optimal_quotes_function_S} shows the optimal bid and ask quote shifts for different values of the price and the mean-reversion parameter $R$. If $R=0,$ then there is no deterministic part in the dynamics of the price, so the strategy is indifferent to the price level. If $R>0$, then the strategy of the market maker has a speculative component; the market maker uses the difference between the price $S$ and its long-term mean $\overline S = 100$ to determine the price of liquidity at the bid and at the ask. More precisely, when the price $S$ is above $\overline S$, the market maker expects the prices to decrease and revert back  to $\overline S$, so she increases the price of liquidity at the bid to maximise her profits from sellers, and she decreases the the price of liquidity at the ask to attract buyers. Similarly, when the price $S$ is below $\overline S$, the market maker expects the prices to increase, so she raises the price of liquidity at the ask and reduces it at the bid.

\begin{figure}[H]
    \centering
    \includegraphics{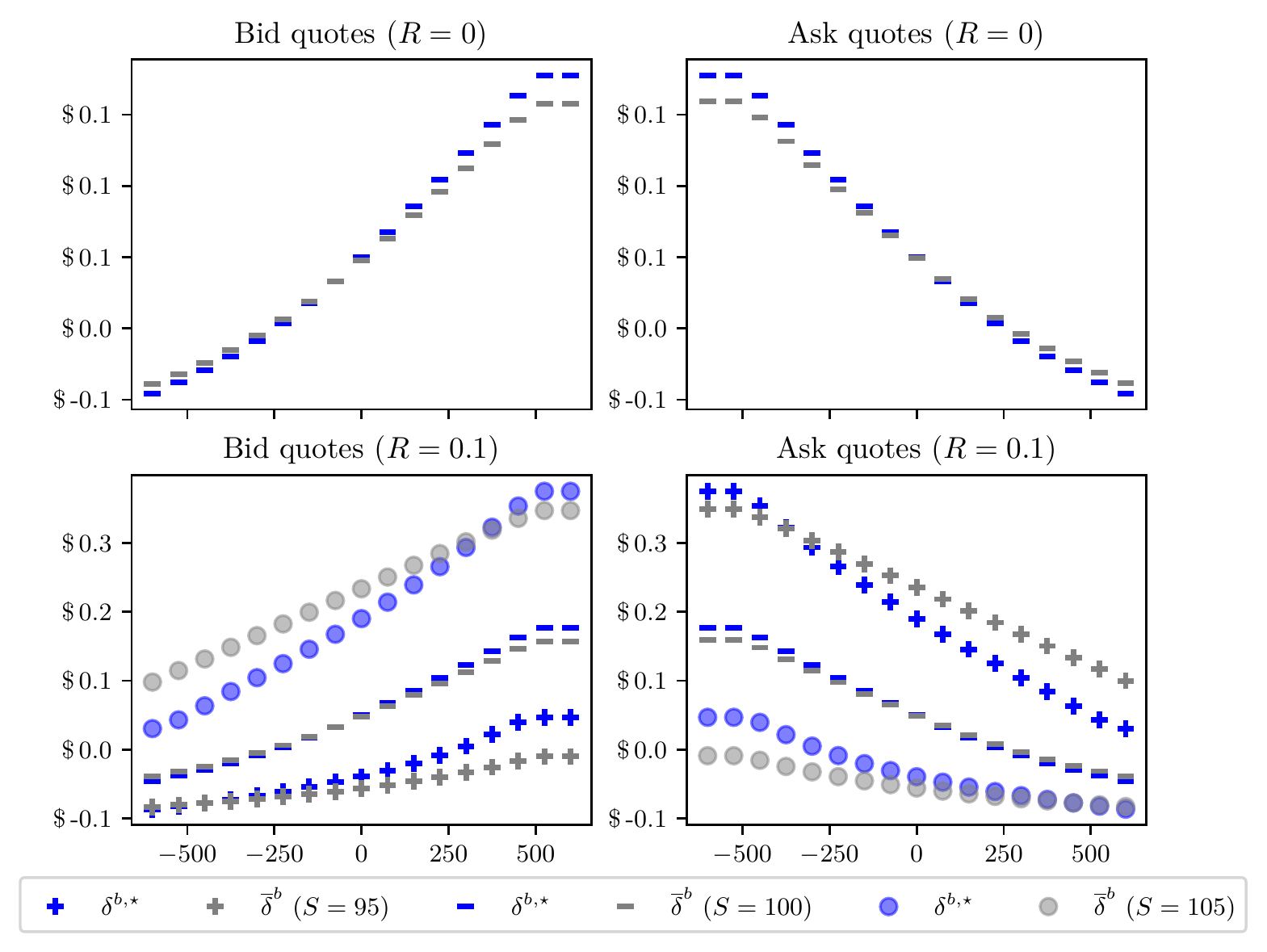}
    \caption{Optimal \eqref{closedloopoptimalcontrols} (in blue) and approximated \eqref{closedloopoptimalcontrols_approx} (in grey) bid and ask shifts as a function of price $S$, mean reversion parameter $R$, and inventory $q$, when the RFQ size is $z=100$ shares.}
    \label{fig:optimal_quotes_approx_1}
\end{figure}

\paragraph{\textbf{Quality of the approximation}} Figure \ref{fig:optimal_quotes_approx_1} shows the optimal shifts  \eqref{closedloopoptimalcontrols} and the approximated shifts \eqref{closedloopoptimalcontrols_approx} for different values of the price $S$, the mean reversion parameter $R\,,$ and the inventory $q$.  When $R=0$ the quality of the approximation is satisfactory for different levels of the inventory; recall that the quoting strategy does not depend on the price level. When $R>0\,$ the approximated bid quotes deviate from the optimal bid quotes when (i) the price $S$ is above $\overline S$ and the inventory of the market maker is large and negative, and when (ii) the price $S$ is below $\overline S$ and the inventory of the market maker is large and positive.  Similarly, the  approximated ask quotes deviate from the optimal ask quotes when (i) the price $S$ is below $\overline S$ and the inventory of the market maker is large and positive, and when (ii) the price $S$ is above $\overline S$ and the inventory of the market maker is large and negative. Most importantly, the approximated quotes capture all the relevant financial effects of the optimal quoting strategy: (i) the price of the liquidity increases with inventory at the bid and decreases with inventory at the ask, and (ii) the price of liquidity increases with the price level at the bid and decreases with the price level at the ask (when $R>0$).

\subsubsection{Multi-asset case} 

We consider a market maker in charge of two cointegrated assets. Table  \ref{table:2DOUParams} shows the parameter values for the multi-OU dynamics of the prices in \eqref{PriceProces} and for the performance criterion in \eqref{eq:mmperfcrit}. In the multi-asset case, the parameter $R \in \mathcal M_2(\mathbbm R)$ is a matrix that drives the deterministic joint dynamics of the prices, and  $\overline S$ is the long-term unconditional mean. The matrix $V$ in Table \ref{table:2DOUParams} indicates that the prices are not correlated, however, the matrix $R$ implies that the space of cointegration vectors is spanned by $(-1, 1)\,,$ i.e., the \emph{cointegration factor} $S^1 - S^2$ is stationary around $0\,.$ 

\begin{table}[H]\footnotesize
\begin{center}
\begin{tabular}{c | c  c  c  c | c  c  c } 
Parameter &  $S_0 = \left(S_0^1, S_0^2\right)$ &  $\overline S = \left(\overline S^1, \overline S^2\right)$ & $R$ & $V$ & $T$ & $\Gamma$ & $\rho$\\
\hline
Value & $\left(\$\,100, \$\,100\right)$ & $\left(\$\,100, \$\,100\right)$  & $\left(\begin{array}{cc}
            0.5 & -0.5\\
            -0.5 & 0.5
            \end{array}\right)$  & 
$\left(\begin{array}{cc}
            1 & 0\\
            0 & 1
            \end{array}\right)$ & $7$ days & $\left(\begin{array}{cc}
            0 & 0\\
            0 & 0
            \end{array}\right)$ & $5\cdot 10^{-3} \, \$^{-1}$ \\[0.5ex]             
\hline
\end{tabular}

\end{center}
\caption {Parameters values for the asset prices and for the performance criterion of the market maker.}
\label{table:2DOUParams}
\end{table}



\begin{figure}[H]
    \centering
    \includegraphics{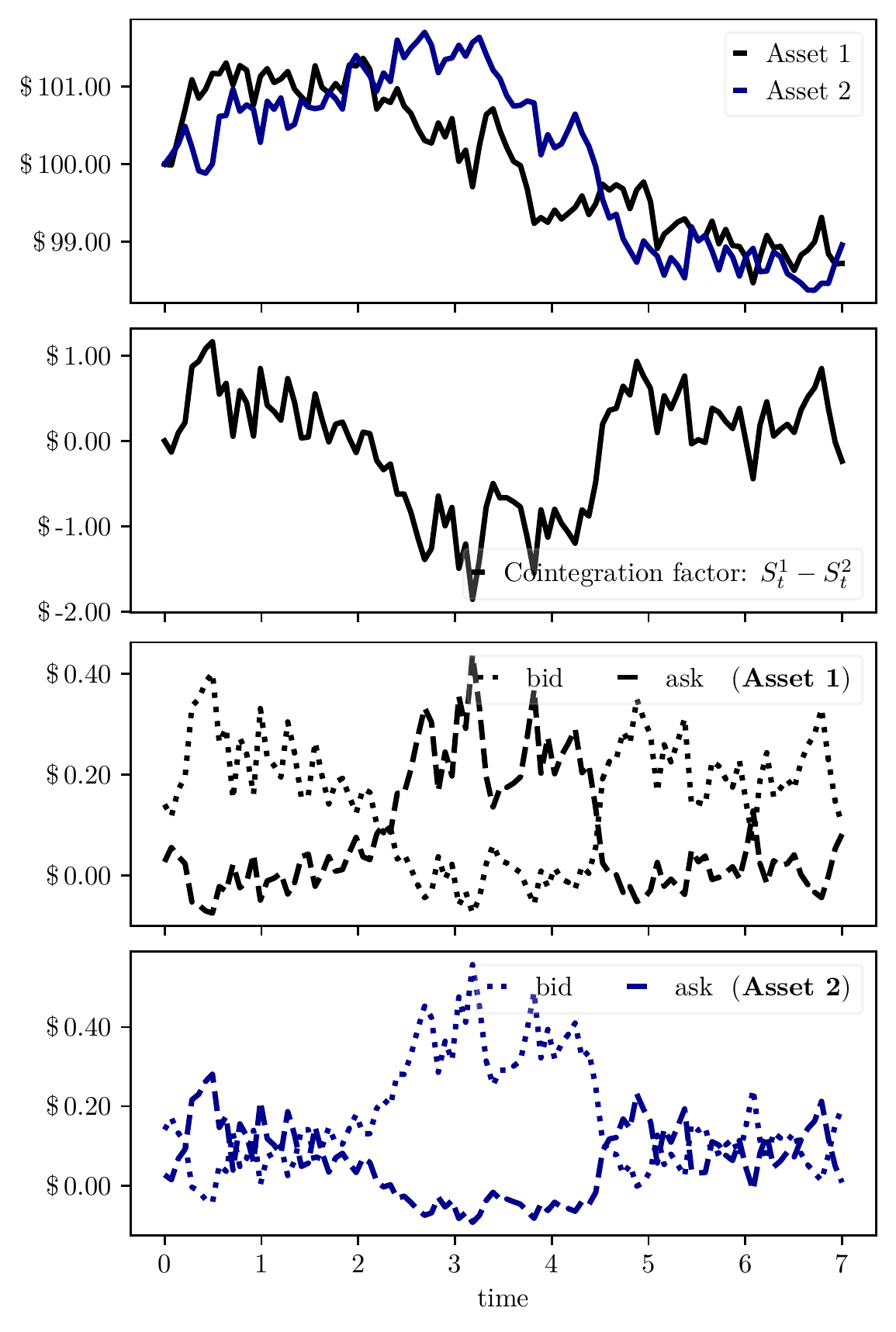}
    \caption{First two panels: Simulation of multi-OU dynamics with the parameter values in Table \ref{table:2DOUParams} and the corresponding cointegration factor $S^1 - S^2$. Last two panels: price of liquidity at the bid and the ask for asset $1$ and asset $2$. The quotes are for an RFQ size $z=100$ and assume the inventory of the market maker is zero.}
    \label{fig:MD_optimal_quotes_cointegration}
\end{figure}

To obtain the optimal quote shifts $\delta^{b,\star}$ and $\delta^{a,\star}$, one should employ a numerical scheme in dimension five; one for time, two for inventory in both assets, and two for the asset prices of both assets, which becomes rapidly intractable with fine grids. In contrast, the approximated shifts \eqref{closedloopoptimalcontrols_approx} are quickly computed with Riccati ODE solvers. Figure \ref{fig:MD_optimal_quotes_cointegration} shows the approximated bid and ask shifts for both assets for a simulation path of both prices. 

The first two panels show that the prices are cointegrated and that the cointegration factor $S^1 - S^2$ is mean reverting to $0\,.$ The last two panels show the price of liquidity at the bid and the ask for both assets as a function of the price levels. In contrast to the single asset case, the speculative component of the quoting strategy uses the cointegration factor $S^1 - S^2$ instead of the prices of the individual assets. In particular, when the cointegration factor is above the long-term mean $0$ , i.e., when $S^1 - S^2>0$, the market maker expects the price $S^1$ of asset $1$ to decrease and the price $S^2$ of asset $2$ to increase. Thus, she increases the price of liquidity at the bid for asset $1$ and  at the ask for asset $2$ to maximise profits. Similarly, when $S^1 - S^2<0$, the market maker expects $S^1$ to increase and $S^2$ to decrease. Thus, the market maker increases the price of liquidity at the ask for asset $1$ and at the bid for asset $2$ to maximise profits.

\section{Optimal execution and statistical arbitrage with Bayesian learning of the drift \label{sec:bayesien}}

In this section, we propose a second application of the existence result of Section \ref{sec:riccati}. Consider an agent who is in charge of a portfolio. The agent is uncertain about the drifts in the prices of the assets in the portfolio, and has a prior Gaussian distribution. The model we introduce combines stochastic optimal control, to obtain an optimal trading schedule, with online learning of the drift from observed prices and signals, to update the estimate of the drift while trading. The strategy we derive uses information from prices and signals to enhance trading performance for execution programmes, or to execute statistical arbitrages based on price predictors.

In our model, the latest estimate at time $t$ of the true value of the drift  is the expectation of the drift conditional on the filtration generated by the prices and signals up until time $t$.  The approach is powerful in two ways. First, it uses online learning to continuously adjust the drift to the information in recent observations of prices and predictive signals. Second, it incorporates the uncertainty on the future value of the drift. 

First, we recall the Bayesian learning approach introduced in \cite{bismuth2019portfolio}. Next, we show that the optimal trading problem reduces to solving a DRE similar to that in \eqref{eq:RiccatiGeneral}.

\subsection{Bayesian learning}


We consider a market with $r \in \mathbb N^\star$ assets, and an agent who uses $k \in \mathbb N$ signals, and let $d = k + r$. We introduce the $d$-dimensional Brownian motion $(W_t)_{t \in  [0,T]} = \left(W_t^1,\dots,W_t^d\right)^\intercal_{t\in[0,T]}$ adapted to the filtration $(\mathcal F(t))_{t \in [0,T]}$. The joint dynamics of the $r$ prices and the $k$ signals are modelled by a $d$-dimensional drifted Bachelier process $\left(S_t \right)_{t \in [0,T]} = \left(S_t^1,\dots,S_t^d\right)^\intercal_{t\in[0,T]}$ with dynamics 
\begin{align}
\label{sect1:dynS}
dS_{t} & =  \mu \,dt+ V\, dW_t\,,
\end{align}
where $\mu \in \mathbb R^d$, $V \in \mathcal M_{d}\left(\mathbb R\right)$, and we denote the variance-covariance matrix by $\Sigma = V\,V^\intercal$.

The drifts $\mu$ are unknown to the agent but she has a Gaussian prior distribution that we denote by $m_\mu$. The drift $\mu$ and the Brownian $(W_t)_{t \in  [0,T]}$ are not observed; however, the prices and the signals are continuously observed by the agent and they convey information about the true value of the drift. We use the Bayesian framework to update the estimation of the drift throughout the execution programme using the latest  information available to the agent.


Let $\mathbb F^S = (\mathcal F^S(t))_{t \in [0,T]}$ be the filtration generated by $(S_t)_{t\in[0,T]}$, which is the filtration that represents the information revealed to the agent. Notice that $(W_t)_{t \in  [0,T]}$ is not $\mathbb F^S$-adapted. We introduce the process $(b_t)_{t\in[0,T]}$ defined as 
\begin{align*}
    b_t=\mathbb{E}\left[\left. \mu \right| \mathcal{F}_t^S \right].
\end{align*}

The process $\left(b_t\right)_{t\in[0,T]}$ is the latest estimated value of the drift $\mu$ given the information in the prices and the signals up to time $t$. Now consider a non-degenerate multivariate Gaussian prior, i.e., 
\begin{equation}
\label{eq:priorgauss_AC} m_\mu(dz)=\frac{1}{(2\pi)^\frac d2|\varPi_0|^\frac 12}\exp
\left(
-\frac12 (z-b_0)^\intercal\varPi_0^{-1}(z-b_0)
\right)dz,
\end{equation}
where $b_0\in \mathbb{R}^d$ and $\varPi_0\in S_d^{++}(\mathbb{R})$. The authors in  \cite{bismuth2019portfolio} show that $\forall t\in[0,T],\quad$ 
\begin{align} \label{eq:formGgaussian}
b_t = \varPi(t)\left( \Sigma^{-1} \left( S-S_0\right) +\varPi_0^{-1}b_0 \right),
\end{align}
where $\varPi(t) = \left(\varPi_0^{-1} + t \, \Sigma^{-1} \right)^{-1}.$ They also show that there exists a Brownian motion $\left(\widehat{W}_{t}\right)_{t\in[0,T]}$ adapted to $\mathbb F^S$ and that has the same correlation structure as that of $\left({W}_{t}\right)_{t\in[0,T]}$, and that   the dynamics of $S$ in \eqref{sect1:dynS} in the appropriate filtration are 
\begin{align}
\label{eq:s_eds-2}
    dS_t =&\, \beta_t \, dt + V \, d\widehat{W}_{t} = \,R\,\left(t\right)\left(\overline{S}-S_{t}\right)\,dt+V \, d\widehat{W}_{t}\,,
\end{align}
where we define \begin{align} \label{sect1:defRSbar}
R\,\left(t\right)=- \varPi\left(t\right)\,\Sigma^{-1} \quad \text{and} \quad \overline{S}=S_{0}-\Sigma\,\varPi_{0}^{-1}\,b_{0}\,.
\end{align}





Note that the dynamics of $S$ in \eqref{eq:s_eds-2} are linear and resemble those of a multi-OU process with a time-dependent mean reversion matrix. This allows us to use the results of Section \ref{sec:riccati} to provide a  solution to the problem of optimal execution with Bayesian learning of the drift.


\subsection{Modelling framework and notations}

An agent is in charge of a portfolio consisting of $r$ assets and she wishes to liquidate her initial inventory, or execute a statistical arbitrage if the initial inventory is zero, within a given period of time $[0,T]$, where $T>0$ is fixed. She holds an $r$-dimensional inventory described by the process $(q_t)_{t \in [0,T]} = \left(q^1_t, \ldots, q^r_t \right)^\intercal _{t \in [0,T]}$ with dynamics 
\begin{align*}
dq_t = v_t\, dt\,,
\end{align*}
where the initial inventory $q_0 \in \mathbb R^r$ is known, and where $(v_t)_{t \in [0,T]}= (v^1_t, \ldots, v^r_t)^\intercal _{t \in [0,T]}$ is the trading speed process for each asset. 

The joint dynamics of prices and signals are in \eqref{eq:s_eds-2}, where $S_0 \in \mathbb R^d$ is given and $R$ and $\overline S$ are in \eqref{sect1:defRSbar}.\footnote{We do not consider the permanent price; as justified in more detail in \cite{bergault2021multi}.}  We define the dynamics for the amount $\left(X_{t}\right)_{t\in[0,T]}$ on the agent's cash account as 
\begin{align}
dX_{t}  =  -v_{t}^\intercal \mathcal J S_{t} dt - v_t^\intercal \eta v_t dt,
\end{align}
where $X_0$ is given, $\eta \in \mathcal S^{++}_r (\mathbb R)$ is the positive-definite and symmetric price impact parameter which models execution costs incurred by the trader, and $\mathcal J$ is an $r \times d$ matrix with $\mathcal J_{ij} = \mathbbm 1_{i=j}\,.$ 

As in the previous section, the agent maximises the performance criterion \eqref{eq:mmperfcrit} and admissible strategies are in the set $\mathcal{A} = \mathcal{A}_0$, where for $t \in [0,T]$, 
\begin{align*}
\mathcal{A}_t =& \left\lbrace (v_s)_{s\in [t,T]},\ \mathbb{R}^r\textrm{-valued, }\ \mathcal{F}^S\text{-adapted},\right. \left.\textrm{and satisfying a linear growth condition w.r.t } (S_s)_{s \in [t,T]}\right\}.
\end{align*} 

\subsection{From HJB to DRE}

The value function $\vartheta:[0,T]\times\mathbb{R}\times\mathbb{R}^r\times\mathbb{R}^d \rightarrow \mathbb R$ associated with the problem solves the HJB 
\begin{align*} 
0 =& \, \partial_{t}w
+\left(\overline{S}-S\right)^\intercal R^\intercal  \nabla_S w +\dfrac{1}{2} \mathrm{Tr} \left( \Sigma D^2_{SS}w \right)  +  \sup_{v\in\mathbb{R}^d}\left\{ v^\intercal \nabla_q w - \left(v^\intercal \mathcal J S +v^\intercal \eta v\right)\partial_{x}w \right\}\,,
\end{align*}
for all $(t,x,q,S) \in [0,T) \times \mathbb R \times \mathbb R^r \times \mathbb R^d$ with terminal condition 
\begin{eqnarray}\label{eq:HJB_liq_fin}
w\left(T,x,q,S\right) &=& - \exp(-\rho(x+q^\intercal \mathcal J S - q^\intercal \Gamma q))\,,\quad  \forall (x,q,S)\in \mathbb{R}\times
\mathbb{R}^r\times
\mathbb{R}^d\,. 
\end{eqnarray}

Similar to the previous section, one consider the two ansatzs  \begin{eqnarray*}
w\left(t,x,q,S\right) & = & - \exp\left(-\rho\left(x+q^\intercal \mathcal J S +\theta(t,q,S)\right)\right)\,, \\
\theta(t,q,S)  & =&  q^\intercal A(t)q + q^\intercal B(t) S + S^\intercal C(t)S + D(t)^\intercal q + E(t)^\intercal S + F(t)\,, 
\end{eqnarray*}
to obtain an ODE system in $A\in C^{1}\left(\left[0,T\right], \mathcal S_r\left(\mathbb{R}\right)\right)$, $B \in C^{1}\left(\left[0,T\right],M_{r,d}(\mathbb{R})\right)$, $C \in C^{1}\left(\left[0,T\right],S_d(\mathbb{R})\right)$, $D \in C^{1}\left(\left[0,T\right],\mathbb{R}^r\right)$, $E \in C^{1}\left(\left[0,T\right],\mathbb{R}^d\right)$, and $F \in C^{1}\left(\left[0,T\right],\mathbb{R}\right)$:
\begin{align}
\label{ODEa}
\begin{cases}
\dot{A}(t) = \frac{\rho}{2} (B(t) + \mathcal J) \Sigma (B(t)^\intercal + \mathcal J) - A(t) \eta^{-1} A(t) \\
\dot{B}(t) = (B(t)+ \mathcal J)R(t) + 2\rho  (B(t) + \mathcal J) \Sigma C(t) - A(t) \eta^{-1 } B(t)\\
\dot{C}(t) = R(t)^\intercal C(t) + C(t)R(t) + 2\rho C(t)\Sigma C(t) - \frac 1{4} B(t)^\intercal \eta^{-1} B(t) \\
\dot{D}(t) = -(B(t)+\mathcal J) R(t)\overline S  + \rho  (B(t) + \mathcal J)\Sigma E(t) - A(t)\eta^{-1}D(t)\\
\dot{E}(t) = -2 C(t)R(t)\overline S + R(t)^\intercal E(t) +2\rho C(t)\Sigma E(t) - \frac 1{2} B(t)^\intercal \eta^{-1} D(t)\\
\dot{F}(t) = -\overline S^\intercal R(t)^\intercal  E(t) - Tr(\Sigma C(t)) +\frac{\rho}2 E(t)^\intercal \Sigma E(t) - \frac 1{4}D(t)^\intercal \eta^{-1} D(t)\,,
\end{cases}
\end{align}
with terminal conditions $A(T) = -\Gamma$ and  $B(T)=  C(T)= D(T)= E(T)= F(T) = 0\,.$ 

The ODE system \eqref{ODEa} can be solved sequentially. First, one solves the subsystem in $\left(A, B, C\right)$, then the subsystem in $\left(D, E\right)$ becomes linear, and finally $F$ is obtained with an integration. Note that the subsystem of ODEs in $\left(A, B, C\right)$ can be written as the matrix DRE  in $P(t) = \begin{pmatrix} A(t) & \frac 12 B(t)^\intercal \\ \frac 12 B(t) & C(t) \end{pmatrix}:$  \begin{align}
\label{eq:LT:riccati}
\dot{P}(t) = Q + Y(t)^\intercal \, P(t) +  P(t)\, Y(t) +  P(t)\,U\,P(t),
\end{align}
with terminal condition $P(T) = \begin{pmatrix} -\Gamma & 0 \\ 0 & 0 \end{pmatrix}$ where  \begin{align*}
Q(t) = & \begin{pmatrix} \frac 12 \rho \Sigma^r & \frac 12 \mathcal J R(t)\\ \frac 12 R(t)^\intercal \mathcal J^\intercal & 0 \end{pmatrix}, & Y(t) = & \begin{pmatrix} 0 & 0\\ \rho \Sigma \mathcal J^\intercal & R(t)\end{pmatrix} , & 
U = & \begin{pmatrix} -\eta^{-1} & 0 \\ 0 & 2\rho \Sigma \end{pmatrix}\,.
\end{align*}

Assumptions \ref{assume:1} and \ref{assume:2} hold, in particular, the matrices $R\left(s\right)$ and $R\left(u\right)$ commute for all $\left(s, u\right) \in [0, T].$ The DRE \eqref{eq:LT:riccati} is similar to the DRE \eqref{PODECOMPREHENSIVE}, which is a special case of the family of problems introduced in Section \ref{sec:riccati}. Thus, Theorems \ref{thm1} and \ref{thm2}  solve the problem of optimal trading with Bayesian learning of the drift, in particular  \eqref{eq:LT:riccati} admits a unique solution. In practice, efficient ODE solving techniques rapidly compute the solution $P$ to the DRE \eqref{eq:LT:riccati}. Finally, we do not study the performance of the optimal strategy which has been thoroughly studied in \cite{bismuth2019portfolio}.

\section{Conclusion}
We obtained an existence and uniqueness result for a type of matrix DREs with indefinite matrix coefficients. The result is used to solve two algorithmic trading problems for liquidity providers and takers. The first problem is multi-asset market making, and we derived an approximation strategy which uses signals to enhance trading performance and a trading venue to hedge risk. The second problem is multi-asset optimal execution and statistical arbitrage, where the agent updates her estimation of the drift throughout the execution programme.

\clearpage
\appendix
\section{Proof of Proposition \ref{prop1} \label{appendixproofprop1}}
Take $\left(t,x,y,z\right) \in \left[0, T\right] \times \mathbb R^d \times \mathbb R^r \times \mathbb R,$ and consider the control $\underline u = \left(0\right)_{s\in[t,T]}.$ The dynamics of the problem starting at $t$ become 
\begin{align*}
\begin{cases}
y_{s}^{t,y,\underline{u}} & =y\\
dz_{s}^{t,x,z,\underline{u}} & =- \left(x_{s}^{t,x}\right)^{\intercal}C\left(s\right)x_{s}^{t,x}ds\\
dx_{s}^{t,x} & =R\,\left(s\right)x_{s}^{t,x}ds+V\left(s\right)dW_{s}\:,
\end{cases}
\end{align*}
and the performance criterion for the strategy $\underline u$ becomes $$\mathbb{E}\left[-\exp\left(-\rho\left(z-\int_{t}^{T}\left(x_{s}^{t,x}\right)^{\intercal}C\left(s\right)x_{s}^{t,x}\,ds+y^{\intercal}\varUpsilon^{\intercal}x_{T}^{t,x}-\left(x_{T}^{t,x}\right)^{\intercal}\varPsi x_{T}^{t,x}-y^{\intercal}\Gamma y\right)\right)\right].$$

In the remainder of the proof, the controlled process notation is dropped, i.e., we write $x_{s} = {x}_{s}^{t,x},$ $y_{s} = y_{s}^{t,y,\underline u},$ and $z_{s} = z_{s}^{t,x,z,\underline u},$ for all $s\in[t,T].$ Next, write the solution of the SDE for the dynamics of $x$ for all $s\in[t,T]$ as 
\begin{align}
\label{eq:proof_solx}
x_{s}=\underline{R}\left(t, s\right)x+\int_{t}^{s}\underline{V}\left(u, s\right)dW_{u},
\end{align}
where 
\begin{align*}
\begin{cases}
\underline{R}\left(t, s\right) & =\exp\left(\int_{t}^{s}R\left(u\right)du\right)\ ,\\
\underline{V}\left(u, s\right) & =\exp\left(\int_{u}^{s}R\left(\pi\right)d\pi\right)V\left(u\right)\ .
\end{cases}
\end{align*}

To prove this point, consider the integrator 
$E_{s}=\exp\left(-\int_{t}^{s}R\left(u\right)du\right)x_{s}\ ,$ and notice that $$dE_{s}=\exp\left(-\int_{t}^{s}R\left(u\right)du\right)V\left(s\right)dW_{s},$$ where Assumption \ref{assume:1}-\ref{assumptions:1:iii} is used.\footnote{In particular, when the matrices $R\left(u\right)$ and $R\left(s\right)$ commute for all $\left(u,s\right) \in [t, T]$, so do the matrices $R\left(u\right)$ and $\int_{t}^{s}R\left(\pi\right)d \pi,$ and so do the matrices $R\left(u\right)$ and $\exp\left(\int_{t}^{s}R\left(\pi\right)d\pi\right)\,.$} Next, integrate both sides to find the dynamics in \eqref{eq:proof_solx}. Notice that the solution is normal and write
$$x_{s}\sim\mathcal{N}\left(\underline{R}\left(s\right)x,\underline{\Sigma}\left(t, s\right)\right),$$ where $$\underline{\Sigma}\left(t,s\right)=\int_{t}^{s}\exp\left(\int_{u}^{s}R\left(\pi\right)d\pi\right)\Sigma\left(u\right)\exp\left(\int_{u}^{s}R\left(\pi\right)^{\intercal}d\pi\right)du,$$ for all $s\in[t,T].$

Use the solution \eqref{eq:proof_solx} to write the following equalities in the case of the control $\underline u:$
\begin{align}
\label{eq:proof_quadxsol}
\begin{cases}
\int_{t}^{T}x_{s}^{\intercal}C\left(s\right)x_{s}dt & =x^{\intercal}\utilde{R}\left(t,T\right)x+2x^{\intercal}\utilde{W}\left(t,T\right)+\utilde{C}\left(t,T\right)\ ,\\
x_{T}\varPsi x_{T} & =x^{\intercal}\underset{\vee}{R}\left(t,T\right)x+2x^{\intercal}\underset{\vee}{W}\left(t,T\right)+\underset{\vee}{C}\left(t,T\right)\, ,
\end{cases}
\end{align}
where  $$\begin{cases}
\utilde{R}\left(t,T\right) & =\int_{t}^{T}\underline{R}\left(t,s\right)^{\intercal}C\left(s\right)\underline{R}\left(t,s\right)ds\ ,\\
\utilde{W}\left(t,T\right) & =\int_{t}^{T}\int_{t}^{s}\underline{R}\left(t,s\right)^{\intercal}C\left(s\right)\underline{V}\left(t,u\right)dW_{u}ds\ ,\\
\utilde{C}\left(t,T\right) & =\int_{t}^{T}\left(\int_{t}^{s}\underline{V}\left(t,u\right)dW_{u}\right)^{\intercal}C\left(s\right)\left(\int_{t}^{s}\underline{V}\left(t,u\right)dW_{u}\right)ds\ ,\\
\underset{\vee}{R}\left(t,T\right) & =\underline{R}\left(t,T\right)^{\intercal}\varPsi\underline{R}\left(t,T\right)\:,\\
\underset{\vee}{W}\left(t,T\right) & =\underline{R}\left(t,T\right)\varPsi\int_{t}^{T}\underline{V}\left(t,s\right)dW_{s}\:,\\
\underset{\vee}{C}\left(t,T\right) & =\left(\int_{t}^{T}\underline{V}\left(t,s\right)dW_{s}\right)^{\intercal}\varPsi\left(\int_{t}^{T}\underline{V}\left(t,s\right)dW_{s}\right)\,.
\end{cases}$$

Use Fubini for $\utilde{W}$ and write  $$\begin{cases}
\utilde{W}\left(t,T\right) & \sim\mathcal{N}\left(0,\utilde{\Sigma}\left(t,T\right)\right)\, ,\\
\underset{\vee}{W}\left(t,T\right) & \sim\mathcal{N}\left(0,\underset{\vee}{\Sigma}\left(t,T\right)\right)\, ,
\end{cases}$$ where $$\begin{cases}
\utilde{\Sigma}\left(t,T\right) & =\int_{t}^{T}\left(\int_{s}^{T}\underline{R}\left(t,s\right)^{\intercal}C\left(s\right)\underline{V}\left(t,u\right)du\right)^{\intercal}\left(\int_{s}^{T}\underline{R}\left(t,s\right)^{\intercal}C\left(s\right)\underline{V}\left(t,u\right)du\right)ds\, ,\\
\underset{\vee}{\Sigma}\left(t,T\right) & =\int_{t}^{T}\underline{R}\left(t,T\right)\varPsi\underline{V}\left(t,s\right)\underline{V}\left(t,s\right)^{\intercal}\varPsi^{\intercal}\underline{R}\left(t,T\right)^{\intercal}ds\, .
\end{cases}$$

Finally, use the definition of the value function in \eqref{eq:vfdef}, equations \eqref{eq:proof_quadxsol}, and Hölder inequality to write \begin{align}\label{eq:prooflb:interm}
&\vartheta(t,x,y,z)\\\geq&\mathbb{E}\left[-\exp\left(-\rho\left(z-y^{\intercal}\Gamma y-\int_{t}^{T}\left(x_{s}^{t,x}\right)^{\intercal}C\left(s\right)x_{s}^{t,x}ds-\left(x_{T}^{t,x}\right)^{\intercal}\varPsi x_{T}^{t,x}-y^{\intercal}\varUpsilon^{\intercal}x_{T}^{t,x}\right)\right)\right]\\=&-\mathbb{E}\left[\exp\left(-\rho\left(-2x^{\intercal}\utilde{W}\left(t,T\right)-\utilde{C}\left(t,T\right)-2x^{\intercal}\underset{\vee}{W}\left(t,T\right)-\underset{\vee}{C}\left(t,T\right)-y^{\intercal}\varUpsilon^{\intercal}x_{T}^{t,x}\right)\right)\right]\\&\exp\left(-\rho\left(z-y^{\intercal}\Gamma y-x^{\intercal}\utilde{R}\left(t,T\right)x-x^{\intercal}\underset{\vee}{R}\left(t,T\right)x\right)\right)\\\geq&-\mathbb{E}\left[\exp\left(8\rho x^{\intercal}\utilde{W}\left(t,T\right)\right)\right]^{1/4}\mathbb{E}\left[\exp\left(8\rho x^{\intercal}\underset{\vee}{W}\left(t,T\right)\right)\right]^{1/4}\\&\mathbb{E}\left[\exp\left(4\rho y^{\intercal}\varUpsilon^{\intercal}x_{T}^{t,x}\right)\right]^{1/4}\mathbb{E}\left[\exp\left(4\rho\utilde{C}\left(t,T\right)+4\rho\underset{\vee}{C}\left(t,T\right)\right)\right]^{1/4}\\&\exp\left(-\rho\left(z-y^{\intercal}\Gamma y-x^{\intercal}\utilde{R}\left(t,T\right)x-x^{\intercal}\underset{\vee}{R}\left(t,T\right)x\right)\right)\\=&-\exp\left(-\rho z+x^{\intercal}\left(8\rho^{2}\utilde{\Sigma}\left(t,T\right)+8\rho^{2}\underset{\vee}{\Sigma}\left(t,T\right)+\rho\utilde{R}\left(t,T\right)+\rho\underset{\vee}{R}\left(t,T\right)\right)x\right)\\&\exp\left(y^{\intercal}\left(2\rho^{2}\varUpsilon^{\intercal}\underline{\Sigma}\left(t,s\right)\varUpsilon+\rho\Gamma\right)y+y^{\intercal}\left(\rho\varUpsilon^{\intercal}\underline{R}\left(t,T\right)\right)x\right)\\&\mathbb{E}\left[\exp\left(4\rho\utilde{C}\left(t,T\right)+4\rho\underset{\vee}{C}\left(t,T\right)\right)\right]^{1/4}.
\end{align}

If Assumption \ref{assume:2} holds, then the term $\mathbb{E}\left[\exp\left(4\rho\utilde{C}\left(t,T\right)+4\rho\underset{\vee}{C}\left(t,T\right)\right)\right]^{1/4}$ in the above inequality is one and thus $\vartheta(t,x,y,z) > - \infty. $ If Assumption \ref{assume:2} does not hold, then there exists a small enough $\overline \rho >0\,,$ such that for all $0<\rho<\overline \rho,$ $\ \vartheta(t,x,y,z) > -\infty.$

\section{Proof of Theorem \ref{thm1} \label{appendixproofthm1}}
\allowdisplaybreaks
\paragraph{\underline{From HJB to Riccati}}
First, the HJB equation associated with the control problem \eqref{eq:perfcriterion0} is
\begin{align}
\label{eq:hjbequationcontrol}
0=&\,\partial_{t}w+\sup_{u}\left\{ u^{\intercal}O\left(t\right)^{\intercal}\nabla_{y}w-u^{\intercal}A\left(t\right)u\partial_{z}w-u^{\intercal}B\left(t\right)x\partial_{z}w\right\} \\&-x^{\intercal}C\left(t\right)x\partial_{z}w+x^{\intercal}R\left(t\right)^{\intercal}\nabla_{x}w+\frac{1}{2}\text{Tr}\left(\Sigma\left(t\right)D_{xx}^{2}w\right),
\end{align}
for all $\left(t,x,y,z\right) \in [0,T)\times\mathbb R^d \times \mathbb R^r \times \mathbb R,$ with terminal condition 
\begin{align}
\label{eq:termcondw}
w\left(T,x,y,z\right)=-\exp\left(-\rho\left(z-\left(\begin{matrix}
x\\
y
\end{matrix}\right)^{\intercal}\left(\begin{matrix}
\varPsi & \frac{1}{2}\varUpsilon\\
\frac{1}{2}\varUpsilon^{\intercal} & \Gamma
\end{matrix}\right)\left(\begin{matrix}
x\\
y
\end{matrix}\right)\right)\right)\,.
\end{align}

Consider the function $\theta$ defined in \eqref{eq:ansatz1} and notice that it solves the HJB 
\begin{align}
\label{eq:hjbthetacontrolproblem0}
0=&\partial_{t}\theta+\sup_{u}\left\{ u^{\intercal}\nabla_{y}\theta-u^{\intercal}A\left(t\right)u-u^{\intercal}B\left(t\right)x\right\} \\&-x^{\intercal}C\left(t\right)x+x^{\intercal}R\left(t\right)^{\intercal}\nabla_{x}\theta+\frac{1}{2}\text{Tr}\left(\Sigma\left(t\right)D_{xx}^{2}\theta\right)-\frac{\rho}{2}\nabla_{x}\theta^{\intercal}\Sigma\nabla_{x}\theta,
\end{align}
with terminal condition 
\begin{align}
\label{eq:termcondtheta}
\theta\left(T,x,y\right) = - \left(\begin{matrix}
x\\
y
\end{matrix}\right)^{\intercal}\left(\begin{matrix}
\varPsi & \frac{1}{2}\varUpsilon\\
\frac{1}{2}\varUpsilon^{\intercal} & \Gamma
\end{matrix}\right)\left(\begin{matrix}
x\\
y
\end{matrix}\right), \quad \forall \left(x,y\right) \in \mathbb R^d \times \mathbb R^r.
\end{align}

The supremum in \eqref{eq:hjbthetacontrolproblem0} is attained for the optimal control 
\begin{align}
\label{eq:ufeedback0}
u_s^{\star}=\frac{1}{2}A\left(s\right)^{-1}\left(\nabla_{y}\theta\left(s,x,y\right)-B\left(s\right)x_{s}\right), \quad \forall s \in [t,T].
\end{align}

Replace in \eqref{eq:hjbthetacontrolproblem0} to obtain the new HJB 
\begin{align}
\label{eq:hjbthetacontrolproblem}
0=&\,\partial_{t}\theta+\frac{1}{4}\left(\nabla_{y}\theta-B\left(t\right)x\right)^{\intercal}A\left(t\right)^{-1}\left(\nabla_{y}\theta-B\left(t\right)x\right)\\&-x^{\intercal}C\left(t\right)x+x^{\intercal}R\left(t\right)^{\intercal}\nabla_{x}\theta+\frac{1}{2}\text{Tr}\left(\Sigma\left(t\right)D_{xx}^{2}\theta\right)-\frac{\rho}{2}\,\nabla_{x}\theta^{\intercal}\Sigma\left(t\right)\nabla_{x}\theta.
\end{align}

Next, use the definition of $\theta$ in \eqref{eq:ansatz1}, replace in the HJB \eqref{eq:hjbthetacontrolproblem}, and identify the terms of the polynomial to obtain the ODE system 
\begin{align}
\label{eq:odesystemproof}
\begin{cases}
\dot{M}_{1}\left(t\right)= & -\frac{1}{4}\left(M_{2}\left(t\right)-B\left(t\right)^{\intercal}\right)A\left(t\right)^{-1}\left(M_{2}\left(t\right)^{\intercal}-B\left(t\right)\right)+C\left(t\right)\\
 & -R\left(t\right)^{\intercal}M_{1}\left(t\right)-M_{1}\left(t\right)R\left(t\right)+2\rho M_{1}\left(t\right)\Sigma\left(t\right)M_{1}\left(t\right),\\
\dot{M}_{2}\left(t\right)= & -\left(M_{2}\left(t\right)-B\left(t\right)^{\intercal}\right)A\left(t\right)^{-1}M_{3}\left(t\right)-R\left(t\right)^{\intercal}M_{2}\left(t\right)+2\rho M_{1}\left(t\right)\Sigma\left(t\right)M_{2}\left(t\right),\\
\dot{M}_{3}\left(t\right)= & -M_{3}\left(t\right)^{\intercal}A\left(t\right)^{-1}M_{3}\left(t\right)+\frac{\rho}{2}M_{2}\left(t\right)^{\intercal}\Sigma\left(t\right)M_{2}\left(t\right),\\
\dot{M}_{4}\left(t\right)= & -\frac{1}{2}\left(M_{2}\left(t\right)+B\left(t\right)^{\intercal}\right)A\left(t\right)^{-1}M_{5}\left(t\right)-R\left(t\right)^{\intercal}\dot{M}_{4}\left(t\right)+2\rho M_{1}\left(t\right)\Sigma\left(t\right)M_{4}\left(t\right),\\
\dot{M}_{5}\left(t\right)= & -M_{3}\left(t\right)^{\intercal}A\left(t\right)^{-1}M_{5}\left(t\right)+\rho M_{2}\left(t\right)^{\intercal}\Sigma\left(t\right)M_{4}\left(t\right),\\
\dot{M}_{6}\left(t\right)= & \frac{1}{4}M_{5}\left(t\right)^{\intercal}A\left(t\right)^{-1}M_{5}\left(t\right)-\text{Tr}\left(\Sigma\left(t\right)M_{1}\left(t\right)\right)+\frac{\rho}{2}M_{4}\left(t\right)^{\intercal}\Sigma\left(t\right)M_{4}\left(t\right),
\end{cases}
\end{align}
with terminal conditions $M_1(T) = -\varPsi\,,$ $M_2(T) = - \varUpsilon,$ $M_3(T) = -\Gamma\,,$  and $M_4(T) = M_5(T) = M_6(T) = 0.$ Note that the ODE system in $\left(M_1, M_2, M_3\right)$ can be solved independently, that the ODE system in $\left(M_4, M_5\right)$ admits the solution $M_4 = M_5 = 0\,,$ and finally that $M_6$ is obtained with a simple integration in \eqref{eq:thm1:M6}.

Theorem \ref{thm1} assumes the DRE \eqref{eq:RiccatiGeneral} with terminal condition \eqref{eq:termcondP} admits a unique solution. Let $M_1 \in C^1 \left([0,T], \mathcal S_d(\mathbb R) \right)$, $M_2 \in C^1 \left([0,T], \mathcal M_{r,d}(\mathbb R) \right)$, and $M_3 \in C^1 \left([0,T], \mathcal S_r(\mathbb R) \right)$ be that solution. Consequently, the ODE system \eqref{eq:odesystemproof} admits a unique solution because $M_4 = M_5 = 0$ and $M_6$ is in \eqref{eq:thm1:M6}, and the candidate optimal control in feedback is 
\begin{align}
\label{eq:ufeedback}
u_{s}^{\star}=\frac{1}{2}A\left(s\right)^{-1}\left(M_{2}\left(s\right)^{\intercal}x_{s}+2M_{3}\left(s\right)y_{s}-B\left(s\right)x_{s}\right), \quad \forall s \in [t,T].
\end{align}


Define  $\theta$ as in \eqref{eq:ansatz1} and notice it solves the HJB equation \eqref{eq:hjbthetacontrolproblem} with terminal condition \eqref{eq:termcondtheta}. Finally, define $w$ as in \eqref{eq:ansatz2} and notice it solves the HJB equation \eqref{eq:hjbequationcontrol} with terminal condition \eqref{eq:termcondw}, which is the HJB associated to our control problem. Thus, $w$ is a candidate solution for the control problem \eqref{eq:vfdef}.

\paragraph{\underline{Admissibility of the optimal control}} To prove that $(u^\star_s)_{s \in [t,T]}$ in \eqref{eq:ufeedback} is well-defined and admissible, consider the Cauchy initial value problem, $\forall s \in [t,T],\quad$  \begin{align}\label{eq:proof:cauchy} \frac{d\tilde{y}_{s}}{ds}=\,\frac{1}{2}A\left(s\right)^{-1}\left(M_{2}\left(s\right)^{\intercal}x_{s}+2M_{3}\left(s\right)y_{s}-B\left(s\right)x_{s}\right)=\,\phi\left(s\right)y_{s}+\Phi\left(s,x_{s}\right)\,,
\end{align}
where $\tilde y_t = y$, $\ \phi\left(s\right)=A\left(s\right)^{-1}M_{3}\left(t\right)$, and $\ \Phi\left(s,x_{s}\right)=\frac{1}{2}A\left(s\right)^{-1}\left(M_{2}\left(s\right)^{\intercal}x_{s}-B\left(s\right)x_{s}\right)\,.$ 

The unique solution to \eqref{eq:proof:cauchy} is  $$\tilde{y}_{s}=\exp\left(\int_{t}^{s}\phi(r)dr\right)\left(y+\int_{t}^{s}\Phi\left(r,x_{r}\right)\exp\left(-\int_{t}^{r}\phi(\epsilon)d\epsilon\right)dr\right)$$ and the optimal control is  $$v_{s}^{\star}=\frac{d\tilde{y}_{s}}{ds}=\phi(s)\exp\left(\int_{t}^{s}\phi(r)dr\right)\left(y+\int_{t}^{s}\Phi\left(r,x_{r}\right)\exp\left(-\int_{t}^{r}\phi(\epsilon)d\epsilon\right)dr\right)+\Phi\left(s,x_{s}\right)\,.$$

Note from the affine form of $\Phi$ in $x$ that $u^\star$ satisfies a linear growth condition in $x$ and is therefore admissible.

\paragraph{\underline{Verification}} Fix $\left(t,x,y,z\right)\in[0,T]\times\mathbb{R}^{d}\times\mathbb{R}^{r}\times\mathbb{R}$ and $=(v_{s})_{s\in[t,T]}\in\mathcal{A}_{t},$. To prove that  $$\mathbb{E}\left[w\left(T,x_{T}^{t,x},y_{T}^{t,u},z_{T}^{t,x,z,u}\right)\right]\leqslant w(t,x,y,z),$$ use Ito's lemma for $s\in[t, T]$ and write  \begin{equation}\label{eq:verifproof:int1}dw_{s}^{u}=\mathcal{L}^{u}w_{s}^{u}ds+\left(\nabla_{x}w_{s}^{u}\right)^{\intercal}V\left(s\right)dW_{s},\end{equation} where the notation  $w(s,x_{s}^{t,x},y_{s}^{t,u},z_{s}^{t,x,z,u})=w_{s}^{u}$ is used and \begin{align*}\mathcal{L}^{u}w_{s}^{u}=&\partial_{t}w_{s}^{u}-\left(x^{\intercal}C\left(s\right)x+u^{\intercal}A\left(s\right)u+x^{\intercal}B\left(s\right)^{\intercal}u\right)\partial_{z}w_{s}^{u}\\&+u^{\intercal}\nabla_{y}w_{s}^{u}+x^{\intercal}R\left(s\right)^{\intercal}\nabla_{x}w_{s}^{u}+\frac{1}{2}\textrm{Tr}\left(\Sigma D_{xx}^{2}w_{s}^{u}\right)\,.\end{align*} Next, use \eqref{eq:ansatz1} and \eqref{eq:ansatz2} to see that   $$\nabla_{x}w_{s}^{u}=-\rho w_{s}^{u}\nabla_{x}\theta_{s}^{u}=-\rho w_{s}^{u}\left(2M_{1}\left(s\right)x_{s}+M_{2}\left(s\right)y_{s}+M_{4}\left(s\right)\right)=w_{s}^{u}\kappa_{s}^{u}\,,$$ where $\kappa_{s}^{u} = -\rho \left(2M_{1}\left(s\right)x_{s}+M_{2}\left(s\right)y_{s}+M_{4}\left(s\right)\right)\,.$ The equality in \eqref{eq:verifproof:int1} thus becomes \begin{equation}\label{eq:verifproof:int2}dw_{s}^{u}=\mathcal{L}^{u}w_{s}^{u}ds+\kappa_{s}^{\intercal}\left(w_{s}^{u}\right)^{\intercal}V\left(s\right)dW_{s}\,.\end{equation}

Next, define \begin{equation}\label{eq:verifproof:int3}\chi_{t,s}^{u}=\exp\left(\int_{t}^{s}{\kappa_{r}^{u}}^{\intercal}V\left(r\right)dW_{r}-\frac{1}{2}\int_{t}^{s}{\kappa_{r}^{u}}^{\intercal}\Sigma\left(r\right)\kappa_{r}^{u}dr\right)\,,\end{equation} and use \eqref{eq:verifproof:int2} and \eqref{eq:verifproof:int3} and Ito's lemma to write \begin{align*}d\left(w_{s}^{u}\left(\chi_{t,s}^{u}\right)^{-1}\right)=&\,dw_{s}^{u}\left(\chi_{t,s}^{u}\right)^{-1}+d\left(\chi_{t,s}^{u}\right)^{-1}w_{s}^{u}-\left(\chi_{t,s}^{u}\right)^{-1}{\kappa_{s}^{u}}^{\intercal}\Sigma\left(s\right)\nabla_{x}w_{s}^{u}ds\\=&\,\left(\chi_{t,s}^{u}\right)^{-1}\left(dw_{s}^{u}-w_{s}^{u}{\kappa_{s}^{u}}^{\intercal}V\left(s\right)dW_{s}+w_{s}^{u}{\kappa_{s}^{u}}^{\intercal}\Sigma\left(s\right)\kappa_{s}^{u}ds-{\kappa_{s}^{u}}^{\intercal}\Sigma\left(s\right)\nabla_{x}w_{s}^{u}ds\right)\\=&\,\left(\chi_{t,s}^{u}\right)^{-1}\left(\mathcal{L}^{u}w_{s}^{u}ds\right)\,. \end{align*}

By definition of $w$, we know that $\mathcal{L}^{u}w_{s}^{u}\le0,$ and that equality holds for the control in \eqref{eq:ufeedback0} which corresponds to the optimal control \eqref{eq:optcontrolsec1}. Thus, $\left(w_{s}^{u}\left(\chi_{t,s}^{u}\right)^{-1}\right)_{s\in[t,T]}$ is a nonincreasing process, consequently  \begin{align*} w\left(T,x_{T}^{t,x},y_{T}^{t,u},z_{T}^{t,x,z,u}\right)\left(\chi_{t,T}^{u}\right)^{-1} \leqslant w(t,x,y,z) &  \implies w\left(T,x_{T}^{t,x},y_{T}^{t,u},z_{T}^{t,x,z,u}\right)\leqslant w(t,x,y,z)\chi_{t,T}^{u}\\ & \implies \mathbb{E}\left[w\left(T,x_{T}^{t,x},y_{T}^{t,u},z_{T}^{t,x,z,u}\right)\right]\leqslant w(t,x,y,z)\mathbb{E}\left[\chi_{t,T}^{u}\right]\,.\end{align*}

The next step is to prove that $\mathbb{E}\left[\chi_{t,T}^{u}\right] = 1.$ Note that $\kappa_s^u$ is affine in $x$ and $y$, and because $\left(y_{s}^{t,u}\right)_{s\in[t,T]}$ satisfies a linear growth condition with respect to $\left(x_{s}^{t,u}\right)_{s\in[t,T]}$, there exists a constant $C^\kappa $ such that, almost surely for all $s<\tau\le T$,  $$\underset{r\in[s,\tau]}{\sup}{\parallel\kappa_{r}^{u}\parallel}^{2}\leqslant C^\kappa\left(1+\underset{r\in[s,\tau]}{\sup}{\parallel W_{r}-W_{s}\parallel}^{2}\right)\,.$$ 

Use the classical properties of the Brownian motion to write  $$\exists \varepsilon >0, \forall s \in [t,T], \hspace{0.3cm} \mathbb E \left[ \exp \left( \frac12 \int_{s}^{ \left(s+\varepsilon \right) \wedge T}  \left(\kappa^{u}_r \right)^\intercal \Sigma \kappa^{u}_r dr \right) \right] < +\infty\,,$$ and use the results in \cite{karatzas2014brownian} to obtain that $(\xi^{u}_{t,s})_{s \in [t,T]}$ is a martingale. Thus, $\mathbb{E}\left[\chi_{t,T}^{u}\right] = 1$ because  $\chi_{t,t}^{u} = 1.$ Finally, conclude that  \begin{align*}\vartheta(t,x,y,z) = \mathbb{E}\left[-\exp\left(-\rho\left(z_{T}^{t,x,z,u^\star}-\left(\begin{matrix}
x_{T}^{t,x}\\
y_{T}^{t,y,u^\star}
\end{matrix}\right)^{\intercal}\left(\begin{matrix}
\varPsi & \frac{1}{2}\varUpsilon\\
\frac{1}{2}\varUpsilon^{\intercal} & \Gamma
\end{matrix}\right)\left(\begin{matrix}
x_{T}^{t,x}\\
y_{T}^{t,y,u^\star}
\end{matrix}\right)\right)\right)\right] = w(t,x,y,z)\,.
\end{align*}


\section{Proof of Theorem \ref{thm2} \label{appendixproofthm2}}


Cauchy-Lipschitz theorem gives existence and uniqueness of a left-maximal solution on the set $\left(T-\xi,T\right]\,.$ Theorem \ref{thm1} ensures that the associated function $w$ defined in \eqref{eq:ansatz1} is the value function of the problem on any interval $[\tau, T],$ where $\tau \in \left(T-\xi,T \right)\,.$

The proof consists in establishing by contradiction that $\xi$ cannot be finite. This is achieved by providing a-priori bounds for a solution $P$ of \eqref{eq:RiccatiGeneral} with terminal condition \eqref{eq:termcondP}. This proves that the solution does not blow up in finite time.\footnote{Bounds are in the sense of the natural order on symmetric matrices: for $\underline M,\overline M  \in \mathcal S_d(\mathbb R)$, $\underline M \leq \overline M $ if and only if $\overline M - \underline M  \in \mathcal S^{+}_d(\mathbb R)$.} The proof uses the original control problem to find bounds for the function $\theta$ in the form of polynomials of degree $2$ in $\left(x,y\right),$ which translate into bounds for the quadratic term of $\theta,$ i.e., $P.$

\paragraph{\underline{Lower bound}} Fix $\tau \in (T-\xi, T),$ take $\left(t,x,y,z\right) \in \left[\tau, T\right] \times \mathbb R^d \times \mathbb R^r \times \mathbb R,$ and consider the control $\underline u = \left(0\right)_{s\in[t,T]}.$ The same arguments as in the proof \ref{appendixproofprop1} lead to the inequality in \eqref{eq:prooflb:interm} which  translates into 
\begin{align*}
\theta\left(t,x,y\right)=&\left(\begin{matrix}
x_{t}\\
y_{t}
\end{matrix}\right)^{\intercal}P\left(t\right)\left(\begin{matrix}
x_{t}\\
y_{t}
\end{matrix}\right)+\left(\begin{matrix}
x_{t}\\
y_{t}
\end{matrix}\right)^{\intercal}\left(\begin{matrix}
M_{4}\left(t\right)\\
M_{5}\left(t\right)
\end{matrix}\right)+M_{6}\left(t\right)\\\geq&\left(\begin{matrix}
x_{t}\\
y_{t}
\end{matrix}\right)^{\intercal}\left(\begin{matrix}
\underline{M}_{1}\left(t\right) & \underline{M}_{2}\left(t\right)\\
\underline{M}_{2}\left(t\right)^{\intercal} & \underline{M}_{3}\left(t\right)
\end{matrix}\right)\left(\begin{matrix}
x_{t}\\
y_{t}
\end{matrix}\right)+\underline{M}_{6}\left(t\right),
\end{align*}
where $$\begin{cases}
\underline{M}_{1}\left(t\right)= & \utilde{R}\left(t,T\right)-2\rho\left(\utilde{\Sigma}\left(t,T\right)+\underset{\vee}{\Sigma}\left(t,T\right)\right)+\underline{R}\left(t,T\right)^{\intercal}\varPsi\underline{R}\left(t,T\right)\,,\\
\underline{M}_{2}\left(t\right)= & -\frac{\rho}{2} \varUpsilon^{\intercal}\underline{R}\left(t,T\right)\,,\\
\underline{M}_{3}\left(t\right)= & -2\rho^{2}\varUpsilon^{\intercal}\underline{\Sigma}\left(t,s\right)\varUpsilon-\rho\Gamma\,,\\
\underline{M}_{6}\left(t\right)= & -\frac{1}{4\rho}\log\left(\mathbb{E}\left[\exp\left(4\rho\utilde{C}\left(t,T\right)+4\rho\underset{\vee}{C}\left(t,T\right)\right)\right]\right)\,.
\end{cases}$$

Note that $\underline{M}_{6}$ exists in both cases of Theorem \ref{thm2}. The above order translates into the same order for the quadratic components of both polynomials of degree $2.$ In particular, one obtains a lower bound for $P$ in the form
\begin{equation}
\label{eq:proof_lb}
\boxed{
P\left(t\right)\ge \left(\begin{matrix}
\underline{M}_{1}\left(t\right) \underline{M}_{2}\left(t\right)\\
\underline{M}_{2}\left(t\right)^{\intercal} \underline{M}_{3}\left(t\right)
\end{matrix}\right).
}
\end{equation}

\paragraph{\underline{Upper bound}} Fix $\tau \in (T-\xi, T),$ take $\left(t,x,y,z\right) \in \left[\tau, T\right] \times \mathbb R^d \times \mathbb R^d \times \mathbb R,$ and notice that  
\begin{align}\label{eq:proofupperbound:ineq}
&\,\mathbb{E}\left[-\exp\left(-\rho\left(z_{T}^{t,x,z,u}-\left(\begin{matrix}
x_{T}^{t,x}\\
y_{T}^{t,y}
\end{matrix}\right)^{\intercal}\left(\begin{matrix}
\varPsi & \frac{1}{2}\varUpsilon\\
\frac{1}{2}\varUpsilon^{\intercal} & \Gamma
\end{matrix}\right)\left(\begin{matrix}
x_{T}^{t,x}\\
y_{T}^{t,y}
\end{matrix}\right)\right)\right)\right]\\\leq&\,\mathbb{E}\left[-\exp\left(-\rho\left(z_{T}^{t,x,z,u}-y_{T}^{t,y}\varUpsilon^{\intercal}x_{T}^{t,x}\right)\right)\right]\\\leq&\,\mathbb{E}\left[-\exp\left(-\rho\left(z-\int_{t}^{T}u_{s}^{\intercal}B\left(s\right)x_{s}^{t,x}ds-y_{T}^{t,y}\varUpsilon^{\intercal}x_{T}^{t,x}\right)\right)\right]\,,
\end{align}
because $A, C, \varPsi, $ and $\Gamma$ are positive matrices.  In the remainder of the proof for the upper bound, the controlled process notation is dropped for readability, i.e., we write $x_{s} = {x}_{s}^{t,x},$ $y_{s} = y_{s}^{t,y, u},$ and $z_{s} = z_{s}^{t,x,z,u},$ for all $s\in[t,T].$ 

Next, use integration by parts to obtain  
$$\int_{t}^{T}u_{s}^{\intercal}B\left(s\right)x_{s}\,ds = y_{T}^{\intercal}B\left(T\right)x_{T}-y^{\intercal}B\left(t\right)x-\int_{t}^{T}y_{s}^{\intercal}\left(B'\left(s\right)+B\left(s\right)R\left(s\right)\right)x_{s}ds-\int_{t}^{T}y_{s}^{\intercal}B\left(s\right)V\left(s\right)dW_{s}\,,$$ and use \eqref{eq:proofupperbound:ineq} to write 
\begin{align}\label{eq:proofupperbound:ineq2}
&\,\sup_{u\in\mathcal{A}_{t}}\mathbb{E}\left[-\exp\left(-\rho\left(z_{T}-\left(\begin{matrix}
x_{T}\\
y_{T}
\end{matrix}\right)^{\intercal}\left(\begin{matrix}
\varPsi & \frac{1}{2}\varUpsilon\\
\frac{1}{2}\varUpsilon^{\intercal} & \Gamma
\end{matrix}\right)\left(\begin{matrix}
x_{T}\\
y_{T}
\end{matrix}\right)\right)\right)\right]\\\leq&\,\exp\left(-\rho\left(z+y^{\intercal}B\left(t\right)x\right)\right)\\&\qquad
\sup_{u\in\mathcal{A}_{t}}\mathbb{E}\left[-\exp\left(-\rho\left(\int_{t}^{T}y_{s}\left(\overline{R}\left(s\right)^{\intercal}x_{s}ds+y_{s}^{\intercal}B\left(s\right)V\left(s\right)dW_{s}\right)-y_{T}^{\intercal}\overline{\varUpsilon}^{\intercal}x_{T}\right)\right)\right]\\\leq&\,\exp\left(-\rho\left(z+y^{\intercal}B\left(t\right)x\right)\right)\sup_{y\in\mathcal{B}_{t}}\mathbb{E}\left[-\exp\left(-\rho\left(\mathcal{V}_{T}-y_{T}^{\intercal}\overline{\varUpsilon}^{\intercal}x_{T}\right)\right)\right]\,,
\end{align}
where $\overline{R}\left(s\right) = B'\left(s\right)+B\left(s\right)R\left(s\right)$, $\ \overline{\varUpsilon} = B\left(T\right)^\intercal+\varUpsilon$, the set $\mathcal B_t$  is
\begin{align}\label{eq:proof:admissiblesetmerton}
\mathcal{B}_t =& \left\lbrace (y_s)_{s\in [t,T]},\ \mathbb{R}^d\textrm{-valued }\ \mathcal{F}^S\text{-adapted, }\right. \left.\textrm{satisfying a linear growth condition w.r.t}\,(x_s)_{s \in [t,T]}\right\}\,,
\end{align} the process $\left(\mathcal V_s\right)_{s\in[t,T]}$ has dynamics $$d\mathcal{V}_{s}=y_{s}^{\intercal}\overline{R}\left(s\right)x_{s}ds+y_{s}^{\intercal}B\left(s\right)V\left(s\right)dW_{s}\,,$$ and the last inequality holds because when  $\left(u_s\right)_{s\in[t,T]} \in \mathcal A_t,$ then $\left(y_s\right)_{s\in[t,T]} \in \mathcal B_t.$ 

Next, to solve the optimisation problem \begin{align} \label{eq:proof:mertonproblem}
\sup_{y\in\mathcal{B}_{t}}\mathbb{E}\left[-\exp\left(-\rho\left(\mathcal{V}_{T}-y_{T}^{\intercal}\overline{\varUpsilon}^{\intercal}x_{T}\right)\right)\right]\,,
\end{align}
define the associated value function  $\overline \vartheta:[0,T]\times\mathbb{R}\times\mathbb{R}^d \rightarrow \mathbb R$ as 
\begin{eqnarray} \label{eq:proof:trading:valuefunc}
\overline \vartheta(t,\mathcal V,x)= \sup_{y\in\mathcal{B}_{t}}\mathbb{E}\left[-\exp\left(-\rho\left(\mathcal{V}_{T}^{t,\mathcal{V},y,x}-y_{T}^{\intercal}\overline{\varUpsilon}^{\intercal}x_{T}^{t,x}\right)\right)\right]]\,.
\end{eqnarray}

The value function $\overline \vartheta$ is the unique solution to the HJB equation  
\begin{align} \label{eq:proof:HJBmerton1}
0=&\,\partial_{t}\overline{w}+x^{\intercal}R\left(t\right)^{\intercal}\nabla_{x}\overline{w}+\frac{1}{2}\text{Tr}\left(\Sigma\left(t\right)D_{xx}^{2}\overline{w}\right)\\&+\sup_{y\in\mathcal{B}}\left\{ x^{\intercal}\overline{R}\left(t\right)^{\intercal}y\partial_{\mathcal{V}}\overline{w}+\frac{1}{2}y^{\intercal}B\left(t\right)\Sigma\left(t\right)B\left(t\right)^{\intercal}y\partial_{\mathcal{V}\mathcal{V}}\overline{w}+\Sigma\left(t\right)B\left(t\right)^{\intercal}y\partial_{\mathcal{V}}\nabla_{x}\overline{w}\right\}\,,
\end{align}
with terminal condition $\overline w(T,\mathcal V,x) = -\exp\left(- \rho \mathcal V\right)$.  Next, use the ansatz $$\overline{w}\left(t,\mathcal{V},x\right)=-\exp\left(-\rho\left(\mathcal{V}+\overline{\theta}\left(t,x\right)\right)\right)$$ to obtain the HJB  \begin{align} \label{eq:proof:HJBmerton2}
0=&\,\partial_{t}\overline{\theta}+x^{\intercal}R\left(t\right)^{\intercal}\nabla_{x}\overline{\theta}+\frac{1}{2}\text{Tr}\left(\Sigma\left(t\right)D_{xx}^{2}\overline{\theta}\right)-\frac{\rho}{2}\nabla_{x}\overline{\theta}^{\intercal}\Sigma\left(t\right)\nabla_{x}\overline{\theta}\\&+\sup_{y\in\mathcal{B}}\{y^{\intercal}\left(\overline{R}\left(t\right)x-\rho B\left(t\right)\Sigma\left(t\right)\nabla_{x}\overline{\theta}\right)-\frac{\rho}{2}y^{\intercal}B\left(t\right)\Sigma\left(t\right)B\left(t\right)^{\intercal}y\}\,,
\end{align}
with terminal condition $\overline \theta(T,x) = 0 $.  So the HJB \eqref{eq:proof:HJBmerton2} becomes 
\begin{align} \label{eq:proof:HJBmerton3}
0=&\,\partial_{t}\overline{\theta}+\frac{1}{2}\text{Tr}\left(\Sigma\left(t\right)D_{xx}^{2}\overline{\theta}\right)+\frac{1}{2\rho}x^{\intercal}\check{R}\left(t\right)x-x^{\intercal}\tilde{R}\left(t\right)\nabla_{x}\overline{\theta}\,,
\end{align}
where  $$\begin{cases}
\check{R}\left(t\right)= & \overline{R}\left(t\right)^{\intercal}\left(B\left(t\right)\Sigma\left(t\right)B\left(t\right)^{\intercal}\right)^{-1}\overline{R}\left(t\right)\\
\tilde{R}\left(t\right)= & -R\left(t\right)^{\intercal}+\overline{R}\left(t\right)^{\intercal}\left(B\left(t\right)\Sigma\left(t\right)B\left(t\right)^{\intercal}\right)^{-1}B\left(t\right)\Sigma\left(t\right)\,,
\end{cases}$$
where Assumption \ref{assume:1}-\ref{assumptions:1:ii} ensures that $B\left(t\right)\Sigma\left(t\right)B\left(t\right)^{\intercal}$ is invertible. The optimal control in feedback form is  \begin{align} \label{eq:proof:feedbackform1} y_{s}^{\star}=&\frac{1}{\rho}\left(B\left(s\right)\Sigma\left(s\right)B\left(s\right)^{\intercal}\right)^{-1}\left(\overline{R}\left(s\right)x-\rho B\left(s\right)\Sigma\left(s\right)\nabla_{x}\overline{\theta}\left(s,x\right)\right)\,.\end{align}

Next, use the ansatz $\overline{\theta}\left(t,x\right)=x^{\intercal}\overline{M}_{1}\left(t\right)x+x^{\intercal}\overline{M}_{4}\left(t\right)+\overline{M}_{6}\left(t\right)$ to obtain the system of ODEs 
\begin{align}\label{eq:proof:ODEsys}
\begin{cases}
\dot{\overline{M}}_{1}\left(t\right)= & -\frac{1}{2\rho}\check{R}\left(t\right)+\tilde{R}\left(t\right)\overline{M}_{1}\left(t\right)+\overline{M}_{1}\left(t\right)\tilde{R}\left(t\right)\\
\dot{\overline{M}}_{4}\left(t\right)= & \tilde{R}\left(t\right)\overline{M}_{4}\left(t\right)\\
\dot{\overline{M}}_{6}\left(t\right)= & -\text{Tr}\left(\Sigma\left(t\right)\overline{M}_{1}\left(t\right)\right)\,,
\end{cases}
\end{align}
with terminal condition $\overline{M}_{1}\left(T\right)=\overline{M}_{4}\left(T\right)=\overline{M}_{6}\left(T\right)=0$. The system of ODEs \eqref{eq:proof:ODEsys} admits a unique solution because the equation in $\overline M_1$ is a linear system. Moreover, $\overline M_4 = 0$ and $\overline M_6$ is obtained by a simple integration. The optimal control in feedback form in \eqref{eq:proof:feedbackform1} becomes \begin{align} \label{eq:proof:feedbackform1} y_{s}^{\star}=&\frac{1}{\rho}\left(B\left(s\right)\Sigma\left(s\right)B\left(s\right)^{\intercal}\right)^{-1}\left(\overline{R}\left(s\right)x-2\rho B\left(s\right)\Sigma\left(s\right)\overline{M}_{1}\left(s\right)x-\rho B\left(s\right)\Sigma\left(s\right)\overline{M}_{4}\left(s\right)\right)\,.\end{align}

Note that we obtain a classical solution to the optimisation problem \eqref{eq:proof:mertonproblem} and a similar verification argument as in Theorem \ref{thm1} ensures that the solution to the HJB corresponds to the value function $\overline \vartheta$. Thus, the inequality \eqref{eq:proofupperbound:ineq2} becomes  $$-\exp\left(-\rho\left(z+\theta\left(t,x,y\right)\right)\right)\leq -\exp\left(-\rho\left(z+y^{\intercal}B\left(t\right)x+\overline{\theta}\left(t,x\right)\right)\right)\,,$$ thus \begin{align*}\theta\left(t,x,y\right)=&\left(\begin{matrix}
x_{t}\\
y_{t}
\end{matrix}\right)^{\intercal}P\left(t\right)\left(\begin{matrix}
x_{t}\\
y_{t}
\end{matrix}\right)+\left(\begin{matrix}
x_{t}\\
y_{t}
\end{matrix}\right)^{\intercal}\left(\begin{matrix}
M_{4}\left(t\right)\\
M_{5}\left(t\right)
\end{matrix}\right)+M_{6}\left(t\right)\\\leq&\left(\begin{matrix}
x_{t}\\
y_{t}
\end{matrix}\right)^{\intercal}\left(\begin{matrix}
\overline{M}_{1}\left(t\right) & \frac{1}{2}B\left(t\right)^{\intercal}\\
\frac{1}{2}B\left(t\right) & 0
\end{matrix}\right)\left(\begin{matrix}
x_{t}\\
y_{t}
\end{matrix}\right)+\overline{M}_{6}\left(t\right)\,. \end{align*}

The above order translates into the same order for the quadratic components of both polynomials of degree $2.$ In particular, one obtains an upper bound for $P$ in the form
\begin{equation}
\label{eq:proof_ub}
\boxed{
P\left(t\right)\le \left(\begin{matrix}
\overline{M}_{1}\left(t\right) \frac{1}{2}B\left(t\right)^{\intercal}\\
\frac{1}{2}B\left(t\right)  0
\end{matrix}\right)\,.
}
\end{equation}

\paragraph{\underline{Existence}} As $\xi$ is considered finite, there exists $\underline M,\overline M \in \mathcal S_d(\mathbb R)$ with $\underline M \le \overline M$ such that $\forall t \in [T - \xi,T]$, $P(t)$ stays in the compact set 
$ \{ M \in \mathcal S_d(\mathbb R)\,  \left|  \underline M\leq M \leq  \overline M \right. \}$. This contradicts the maximality of the solution, hence $\xi = \infty$, proving the result.

\bibliographystyle{elsarticle-harv}
\bibliography{biblio}

\end{document}